\documentclass[aps,prb,10pt,twocolumn,superscriptaddress,nolongbibliography]{revtex4-2}

\usepackage{amsfonts}	
\usepackage{amsmath}    
\usepackage{amssymb}
\usepackage{graphicx}   
\usepackage{verbatim}   
\usepackage{color}      
\usepackage{subfigure}	
\usepackage{hyperref}   
\usepackage{mathtools}

\usepackage{xcolor}

\newcommand{\lef}{\left(}
\newcommand{\rig}{\right)}

\newcommand{\imu}{\mathrm i}

\newcommand{\e}{\mathrm e}

\newcommand{\vF}{v_\mathrm{F}}

\newcommand{\vecA}{\mathbf A}

\newcommand{\vecB}{\mathbf B}
\newcommand{\vece}{\mathbf e}

\newcommand{\veck}{\mathbf k}

\newcommand{\vecn}{\mathbf n}
\newcommand{\vecp}{\mathbf p}

\newcommand{\vecsig}{\boldsymbol \sigma}

\newcommand{\surface}{{\mathrm{surf.}}}
\newcommand{\crit}{{\mathrm{crit.}}}
\newcommand{\Zhang}{{\mathrm{3D}}}
\newcommand{\lattice}{{\mathrm{lat.}}}

\newcommand{\cyl}{{\mathrm{cyl.}}}

\newcommand{\sgn}{{\mathrm{sgn}}}
\newcommand{\kink}{{\mathrm{K}}}

\begin{document}

\title{Magnetotransport signatures of three-dimensional topological insulator nanostructures}

\author{Kristof Moors}
\email[Electronic address: ]{kristof.moors@uni.lu}
\affiliation{University of Luxembourg, Physics and Materials Science Research Unit, Avenue de la Fa\"iencerie 162a, L-1511 Luxembourg, Luxembourg}
\author{Peter Sch\"uffelgen}
\affiliation{Peter Gr\"unberg Institut, Forschungszentrum J\"ulich \& JARA J\"ulich-Aachen Research Alliance, D-52425 J\"ulich, Germany}
\author{Daniel Rosenbach}
\affiliation{Peter Gr\"unberg Institut, Forschungszentrum J\"ulich \& JARA J\"ulich-Aachen Research Alliance, D-52425 J\"ulich, Germany}
\author{Tobias Schmitt}
\affiliation{Peter Gr\"unberg Institut, Forschungszentrum J\"ulich \& JARA J\"ulich-Aachen Research Alliance, D-52425 J\"ulich, Germany}
\author{Thomas Sch\"apers}
\affiliation{Peter Gr\"unberg Institut, Forschungszentrum J\"ulich \& JARA J\"ulich-Aachen Research Alliance, D-52425 J\"ulich, Germany}
\author{Thomas L. Schmidt}
\affiliation{University of Luxembourg, Physics and Materials Science Research Unit, Avenue de la Fa\"iencerie 162a, L-1511 Luxembourg, Luxembourg}


\date{\today}

\begin{abstract}
	We study the magnetotransport properties of patterned 3D topological insulator nanostructures with several leads, such as kinks or Y-junctions, near the Dirac point with analytical as well as numerical techniques.
	The interplay of the nanostructure geometry, the external magnetic field and the spin-momentum locking of the topological surface states lead to a richer magnetoconductance phenomenology as compared to straight nanowires.
	Similar to straight wires, a quantized conductance with perfect transmission across the nanostructure can be realized across a kink when the input and output channels are pierced by a half-integer magnetic flux quantum. Unlike for straight wires, there is an additional requirement depending on the orientation of the external magnetic field. A right-angle kink shows a unique $\pi$-periodic magnetoconductance signature as a function of the in-plane angle of the magnetic field.
	For a Y-junction, the transmission can be perfectly steered to either of the two possible output legs by a proper alignment of the external magnetic field.
	These magnetotransport signatures offer new ways to explore topological surface states and could be relevant for quantum transport experiments on nanostructures which can be realized with existing fabrication methods.
\end{abstract}

\maketitle

\section{Introduction}
A decade ago, three-dimensional topological insulators (3D TIs) entered the scene of condensed matter physics and since then they have remained in the center of attention. Rightfully so, as they offer an interesting theoretical and experimental playground for fundamental research as well as applications, combining relativistic and quantum physics in a single condensed matter system, based on aspects of topology~\cite{Konig2008, Hasan2010, qi2011topological, shen2012topological, franz2013topological}.

Typical properties of 3D TIs are strong spin-orbit coupling, leading to a band inversion in the bulk spectrum, and the appearance of gapless surface states which are protected by time-reversal symmetry. These surface states are well described by a single 2D spin-momentum locked Rashba-Dirac cone, which has been confirmed by angle-resolved photoemission spectroscopy (ARPES) measurements in a wide range of 3D TI materials.

As the 3D TI materials are typically heavily doped, identifying surface state transport in bulk samples has proven to be quite a challenge~\cite{Culcer2012, Bardarson2013}. By studying 3D TI nanostructures instead~\cite{Kong2010, Hong2012, Cha2013}, the surface-to-volume ratio is increased, which in turn increases the detectability of surface state transport. However, confinement generally induces a gap in the surface state spectrum, which increases as the cross section is reduced. Interestingly, a gapless spectrum can be restored through the Aharonov-Bohm (AB) effect by piercing the nanostructure with a half-integer magnetic flux~\cite{Ran2008, Rosenberg2010}.
This leads to unique magnetotransport signatures that one is able to measure systematically in various 3D TI nanowire samples~\cite{Peng2010, Xiu2011, Tang2011, Tian2013, Ning2013, Dufouleur2013, Hamdou2013A, Hamdou2013B, Safdar2013, Hong2014, Cho2015, Baessler2015, Jauregui2016, Sacksteder2016, Kim2016, Arango2016, Gooth2016}, e.g., shifted Shubnikov-de Haas and flux quantum-periodic AB oscillations and weak antilocalization due to the absence of backscattering.
Furthermore, these surface states have been probed directly, using, e.g., nano-ARPES or Kelvin probe microscopy techniques~\cite{Krieg2016, Arango2016, MunozRojo2016}.

There is a solid understanding of uniaxial 3D TI nanowire (or ribbon) surface states in the presence of a magnetic field and different approaches have been introduced to model this system, e.g., effective and continuous (Dirac-like) surface models~\cite{Zhang2009, Ostrovsky2010, Liu2010, Brey2014}, surface and bulk lattice models~\cite{Fu2007, Zhou2017} and a Luttinger liquid description in the 1D limit~\cite{Egger2010}.
Here, we extend these efforts to other 3D TI nanostructures, such as kinks or junctions, which have been proposed as the basic building blocks of 3D TI nanowire circuits for Majorana-based quantum information processing, for example~\cite{Cook2012, Manousakis2017}.

The manuscript is structured as follows. Section~\ref{sec:exp_feasibility} contains a discussion on the experimental feasibility of these structures, as well as concrete fabrication steps. In Sec.~\ref{sec:NW_model}, the models that are used to model the topological surface states and their transport properties are introduced, followed by an overview of the magnetotransport properties in Sec.~\ref{sec:transport}. We conclude and provide an outlook in Sec.~\ref{sec:conclusion}.

\section{Experimental feasibility} \label{sec:exp_feasibility}
We briefly discuss the experimental feasibility of advanced 3D TI nanostructures like Y-junctions, as being considered below. An \emph{in-situ} fabrication method based on molecular beam epitaxy (MBE) grown tetradymite 3D TIs will thus be presented. The necessity of an \emph{in-situ} process is based on observations of a shift of the Fermi level at ambient conditions~\cite{Hoefer2014}.

\begin{figure*}[htb]
	\centering
	\hspace{0.00\linewidth}
	\subfigure[\ ]{\includegraphics[height=0.165\linewidth]{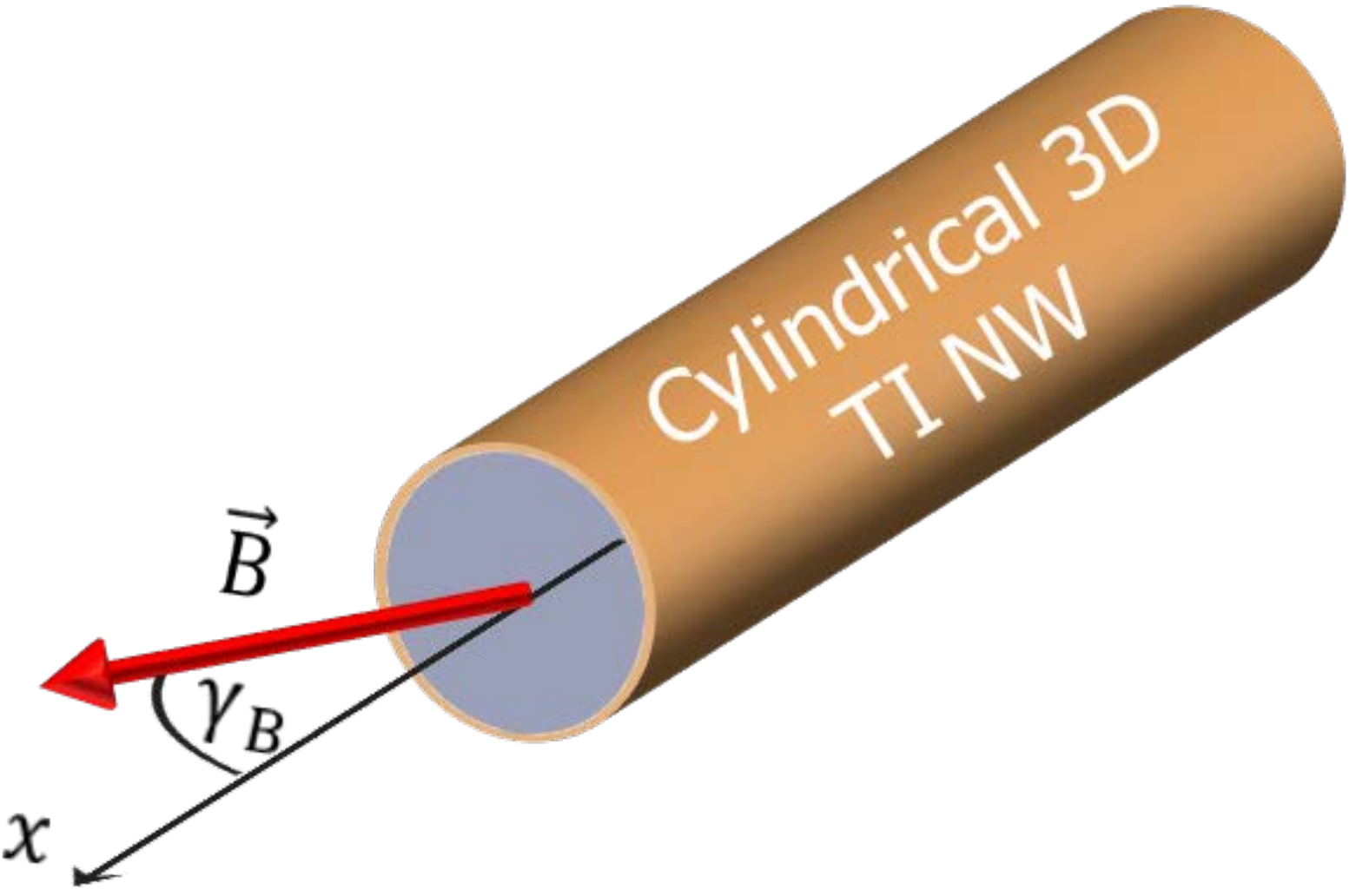}}
	\hspace{0.045\linewidth}
	\addtocounter{subfigure}{1}
	\subfigure[\ ]{\includegraphics[height=0.165\linewidth]{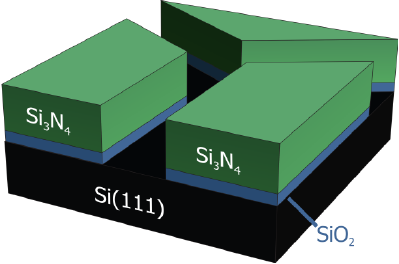}}
	\hspace{0.05\linewidth}
	\addtocounter{subfigure}{1}
	\subfigure[\ ]{\includegraphics[height=0.165\linewidth]{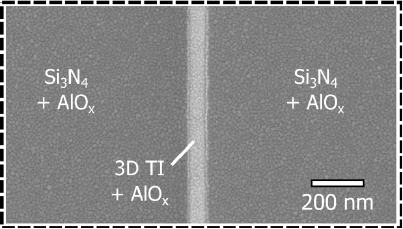}}
	\hspace{0.00\linewidth}
	\newline
	\addtocounter{subfigure}{-4}
	\hspace{0.00\linewidth}
	\subfigure[\ ]{\includegraphics[height=0.165\linewidth]{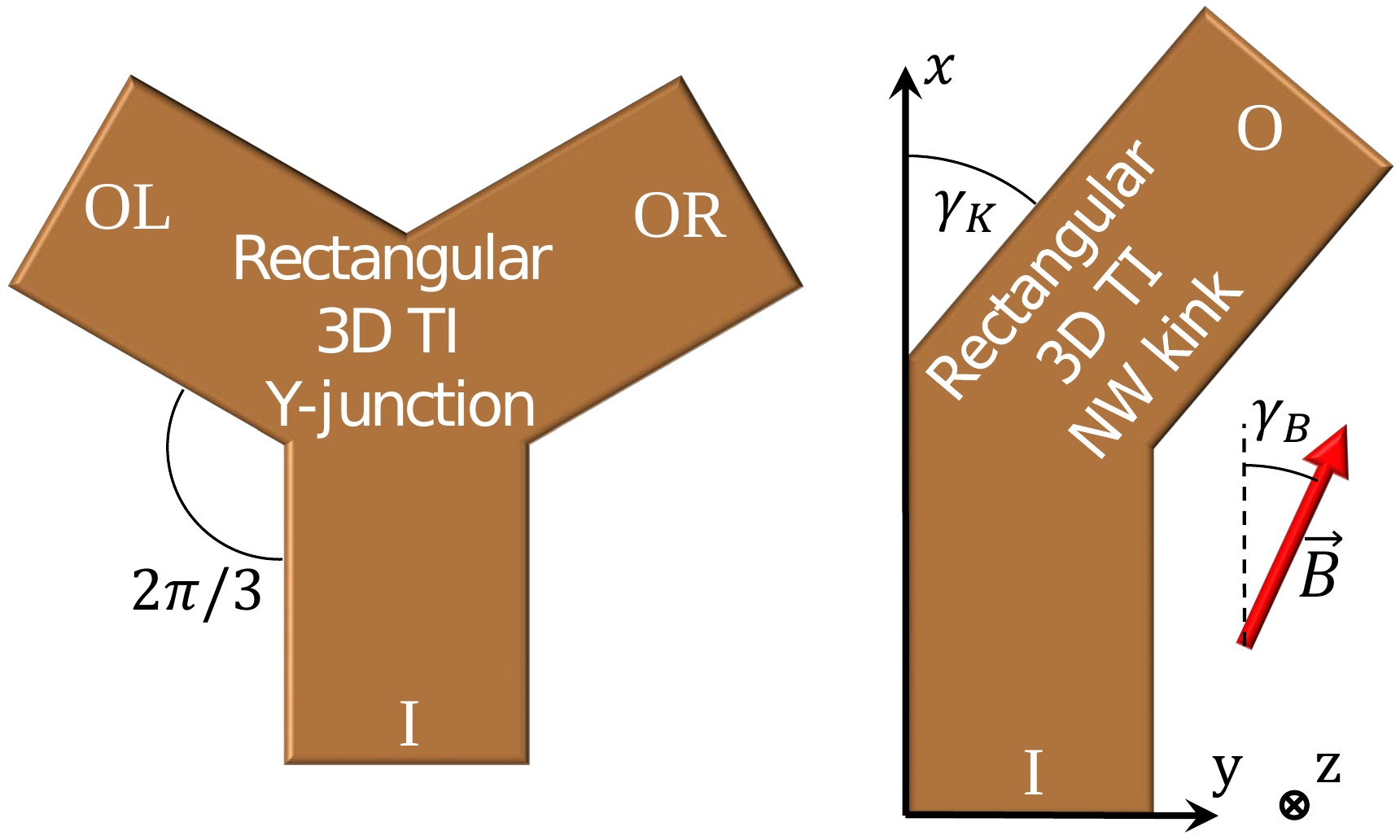}}
	\hspace{0.00\linewidth}
	\addtocounter{subfigure}{1}
	\subfigure[\ ]{\includegraphics[height=0.165\linewidth]{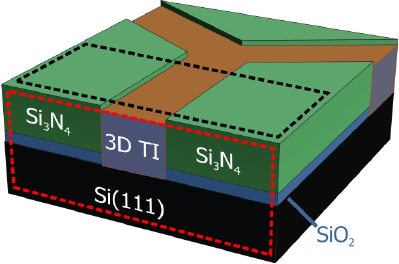}}
	\hspace{0.05\linewidth}
	\addtocounter{subfigure}{1}
	\subfigure[\ ]{\includegraphics[height=0.165\linewidth]{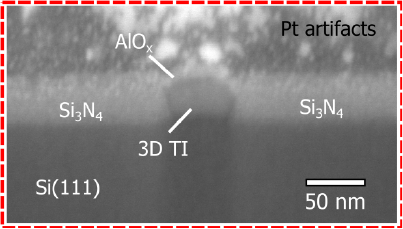}}
	\hspace{0.00\linewidth}
	\caption{
		(a) A sketch of a cylindrical nanowire (NW) in the presence of a magnetic field. The bulk region and its surface, which hosts the topological surface states, are indicated in blue and orange, respectively.
		(b) Top view of rectangular 3D TI nanostructures (kink and Y-junction).
		(c)-(f) Selective area growth of 3D TI nanostructures. (c) A Si(111) substrate with a Si${}_3$N${}_4$/SiO${}_2$ mask layer is prestructured. (d) During MBE growth, the substrate temperature is adjusted such that the 3D TI only nucleates on the Si(111) surface. The surface can be passivated with a final Al capping in-situ. This capping oxidizes at ambient conditions and thereby forms a protective capping layer. (e) SEM top view on a rectangular 3D TI (Bi${}_{0.07}$Sb${}_{0.93}$)${}_2$Te${}_3$ nanowire with a width of 50 nm. The cross section can be seen in (f), prepared by focused ion beam.
	}
	\label{fig:sketches}
\end{figure*}

For fabrication, this requirement excludes etching of 3D TI thin films to nanostructures. Instead, a selective area growth approach~\cite{Kampmeier2016} is employed and improved such that it allows for both the fabrication of advanced 3D TI nanostructures as well as the protection of the Dirac system of the 3D TI. For this technique, a Si(111) substrate with a sacrificial SiO${}_2$ (5~nm) and a Si${}_3$N${}_4$ mask layer (25~nm) is used. After the mask layer has been pre-structured by electron beam lithography and reactive ion etching, the sacrificial layer is smoothly etched in the structures of the mask by hydrofluorid acid in order to uncover the Si(111) substrate in these areas [see Fig.~\ref{fig:sketches}(c)].
During MBE growth, the substrate temperature is set such that nucleation of 3D TI only occurs on the Si(111) surfaces in the structures but not on the Si${}_3$N${}_4$ mask layer [see Fig.~\ref{fig:sketches}(d)]. After selective area growth, the sample is passivated \emph{in situ} with an Al capping layer which oxidizes at ambient conditions and protects the top surface from degradation~\cite{Lang2012}.

In that way, 3D TIs can be shaped in any structure which can be patterned in the mask layer and still preserve their quality. With this process, straight 3D TI nanotrenches with a width down to 50~nm have already been prepared. Scanning electron microscopy (SEM) images of such structures can be seen in Fig.~\ref{fig:sketches}(e) (top view) and in Fig.~\ref{fig:sketches}(f) (cross section). As it has been reported before, the Fermi level in tetradymite 3D TIs can be adjusted by either employing a vertical topological $p$-$n$ junction~\cite{Eschbach2015} or a ternary compound~\cite{Kellner2015}.
To protect the 3D TI surface from ambient conditions even in the area of the electrodes, an additional in-situ stencil lithography method can be used to contact the 3D TI nanostructures~\cite{Schuffelgen2017}.

\section{Model} \label{sec:NW_model}
For straight cylindrical 3D TI nanowires, one can safely resort to an effective 2D surface Rashba-Dirac model. For 3D TI nanostructures with arbitrary shapes and cross sections made of a certain TI material, we employ a tight-binding model based on the effective 3D continuous Hamiltonian, introduced by Zhang \emph{et al}~\cite{Zhang2009}. These two approaches will be discussed in the following subsections.

\subsection{2D Rashba-Dirac Hamiltonian} \label{subsec:NW_cyl}
The surface states of a 3D TI can be effectively described by the bound states of a massive 3D Dirac Hamiltonian for which the Dirac mass undergoes a sign flip at the TI surface. For a flat surface of a 3D TI slab, for example, this leads to the well-known 2D Rashba-Dirac Hamiltonian, $\hat{\mathcal{H}}^\mathrm{RD} = \vF (\hat{p}_y \sigma_x  - \hat{p}_x \sigma_y)$, featuring a single orthogonally spin-momentum locked Dirac cone. For a general curved surface, the 2D Hamiltonian obtains a curvature term and can be written as follows~\cite{Ostrovsky2010}:
\begin{equation} \label{eq:3D_TI_surface}
	\hat{\mathcal{H}}_\surface = -(\vF/2) \left[ \hbar \nabla \cdot \vecn + \vecn \cdot \lef \hat{\vecp} \times \vecsig \rig + \lef \hat{\vecp} \times \vecsig \rig \cdot \vecn \right],
\end{equation}
with $\hat{\vecp}$ the momentum operator on the surface and $\vecn$ a unit vector normal to the surface. We can also add a magnetic field to the surface Hamiltonian through minimal coupling, $\hat{\vecp} \rightarrow \hat{\vecp} - e \vecA$, with electron charge $e$ ($<0$). We will now consider the surface of a cylindrical 3D TI nanowire oriented along the $x$~direction ($\vecn = \vece_r$) with radius $R$ (constant curvature $1/R$) and a constant magnetic field parallel to the cylinder described by the vector potential $\vecA(r, \phi) = B_\parallel r \, \vece_\phi / 2$. We will only consider divergenceless vector potentials throughout the text according to the Coulomb gauge. The eigenstates $\Psi(x, \phi)$ of this system have the following form, based on the symmetries of the system:
\begin{equation} \label{eq:3D_TI_surface_cyl_WF}
	\Psi(x, \phi) \equiv \e^{\imu k x + \imu l \phi}\lef \begin{matrix} \Psi_1 \\ \e^{\imu \phi} \, \Psi_2 \end{matrix} \rig \quad (l = 0, \pm 1, \pm 2 , \ldots),
\end{equation}
with wave vector $k$ and quantized angular momentum $l$.
The Hamiltonian and energy dispersion relation become
\begin{equation} \label{eq:3D_TI_surface_cyl_en}
	\begin{split}
		\hat{\mathcal{H}}_\cyl(j, k) &= \hbar \vF \lef \begin{matrix}
				- j/R & - \imu k \\
				\imu k & j/R
			\end{matrix} \rig, \\
		E_\nu(j, k) &= \nu \, \hbar \vF \sqrt{k^2 + (j/R)^2} \qquad (\nu = \pm 1),
	\end{split}
\end{equation}
with $j \equiv l + 1/2 + \Phi/\Phi_0$ and $\Phi = B_\parallel \pi R^2$ the magnetic flux piercing the cylinder, $\Phi_0 = -h/e > 0$ being the magnetic flux quantum and $j$ the generalized angular momentum, containing contributions from the curvature of the surface and the magnetic flux. The spectrum has a pair of gapless helical subbands ($\nu = \pm 1$, $j = 0$) when a half-integer magnetic flux, $\Phi/\Phi_0 = (2m + 1)/2$ ($m \in \mathbb{Z}$), is piercing the wire.

From Eqs.~(\ref{eq:3D_TI_surface_cyl_WF}) and (\ref{eq:3D_TI_surface_cyl_en}) and box normalization of the wave function ($x \in [-L/2, L/2]$, $k = 2\pi n/L$, $n \in \mathbb{Z}$), we get the following spinor solutions:
\begin{equation} \label{eq:TI_NW_Cyl_EigenV}
	\lef \begin{matrix} \Psi_1 \\ \Psi_2 \end{matrix} \rig = \frac{1}{\sqrt{2 \pi R L}} \left\{ \begin{matrix}
			\lef \begin{matrix} \sin \gamma_\nu(j, k) \\ \imu \cos \gamma_\nu(j, k) \end{matrix} \rig \quad (\nu j \geq 0) \\
			\lef \begin{matrix} \cos \gamma_\nu(j, k) \\ \imu \sin \gamma_\nu(j, k) \end{matrix} \rig \quad (\nu j \leq 0)
		\end{matrix} \right. ,
\end{equation}
with
\begin{equation} \label{eq:TI_NW_Cyl_Angles}
	\gamma_\nu(j, k) \equiv \arctan \lef \frac{\nu k R}{|j| + \sqrt{k^2 R^2 + j^2}} \rig,
\end{equation}
always chosen to lie in the interval $[ -\pi/4, \pi/4 ]$.
This solution is a two-component spinor $\Psi$ that lives on the 2D surface of the cylinder. The corresponding 3D four-component spinor $\chi$ that extends into the bulk region, with constant Dirac mass $M$ (and infinite Dirac mass with opposite sign considered outside of the cylinder), is given by
\begin{equation} \label{eq:3D_TI_surface_3D}
	\chi(x, r, \phi) = \vartheta(R - r) \, e^{|M \vF| (r - R)/\hbar} \lef \begin{matrix} \Psi(x, \phi) \\ \imu \sigma_r \Psi(x, \phi) \end{matrix} \rig,
\end{equation}
allowing us to assign the penetration depth $\lambda = \hbar / | M \vF|$ to the surface state. These four-component spinors can be compared with the four-orbital surface state wave functions that are obtained from the 3D effective model below.

A magnetic field perpendicular to the axial direction of the wire will, in general break, time-reversal symmetry as well as the rotational symmetry around the $x$~axis. To assess its impact, we apply perturbation theory with perturbation Hamiltonian $\hat{\mathcal{H}}_\perp = - e \vF B_\perp R \sin\phi \, \hat{\sigma}_\phi$, arising from a vector potential $\vecA = B_\perp r \sin \phi \, \vece_y$ that corresponds to a magnetic field along the $y$~direction,
$B_\perp \vece_y = B_\perp (\cos \phi \, \vece_r - \sin \phi \, \vece_\phi)$. The wave vector $k$ along the axial (transport) direction remains a valid quantum number, while states with different values for $\nu$ and $j$ get mixed.
The gapless $j=0$ subband remains gapless up to second order in the perpendicular magnetic field $B_\perp$, such that the topological protection is, at least up to a certain extent, maintained in the presence of a perpendicular magnetic field. The first-order correction cancels out completely for the $j=0$ subband while the second-order correction yields a renormalization of the Fermi velocity,
\begin{equation}
	\begin{split}
		\vF \rightarrow \, &\vF \left[ 1 - (e B_\perp R^2)^2/(\pi \hbar)^2 \right] \\
		&= \vF \left[ 1 - (2 B_\perp R^2/\Phi_0)^2 \right],
	\end{split}
\end{equation}
which is symmetric around $E = 0$ and always reduces the magnitude. Unlike for the surface of a 3D TI slab, the energy spectrum remains gapless and the conductance near the Dirac point is unaffected. When the perpendicular magnetic field approaches the critical value of $B_\crit \equiv \Phi_0/(2 R^2)$, the Fermi velocity is renormalized to zero and the perturbative result breaks down. The flat gapless subband spectrum that is obtained in this limit is in agreement with the formation of Landau levels when a strong perpendicular magnetic field is applied~\cite{Shi2014,DeJuan2014,Xypakis2017}. Note that this calculation depends on the rotational symmetry of the nanowire and electron-hole symmetry. To what extent this result holds for general nanostructures with multiple leads will be verified numerically with the 3D effective model presented below.

\subsection{3D effective Hamiltonian} \label{subsec:NW_structure}
For a more realistic (low energy) description of the surface states of various 3D TI materials, Zhang \emph{et al}.\ introduced the following effective 3D continuous Hamiltonian~\cite{Zhang2009, Liu2010}:
\begin{widetext}
\begin{equation} \label{eq:ham_Zhang}
	\begin{split}
		\mathcal{H}^\Zhang (\veck) &\equiv \epsilon(\veck) + \tau_z M(\veck) + \tau_x A_\perp (\sigma_x k_x + \sigma_y k_y) + \sigma_z \tau_x A_z k_z, \\
		\epsilon(\veck) &\equiv C_0 - C_\perp (k_x^2 + k_y^2) - C_z k_z^2, \quad
			M(\veck) \equiv M_0 - M_\perp(k_x^2 + k_y^2) - M_z k_z^2,
	\end{split}
\end{equation}
%
with $z$ the direction of uniaxial anisotropy and $\veck \equiv (k_x, k_y, k_z)$. The four orbitals refer to the electron and hole bands with spin up and spin down ($\mid \! E, \uparrow \rangle$, $\mid \! H, \uparrow \rangle$, $\mid \! E, \downarrow \rangle$ and $\mid \! H, \downarrow \rangle$, respectively), with $\sigma$ ($\tau$) acting on the spin(electron-hole)-subspace.
This Hamiltonian describes an insulator when $C_{\perp, z}^2 < M_{\perp, z}^2$ and a topologically nontrivial regime can be unambiguously assigned, namely when the band inversion parameters $M_\perp$, $M_z$, and the mass (bulk gap) parameter $M_0$ have equal signs: $M_0 M_{\perp, z} > 0$. The band inversion of the E and H bands is governed by $M_{\perp, z}$, while electron-hole asymmetry is captured by $C_{\perp, z}$.
The parameters $A_{\perp, z}$ determine the group velocity of the gapless surface states and finite values for $M_\perp$, $M_z$ prevent the fermion doubling theorem from applying~\cite{nielsen1981absence}. Hence, this Hamiltonian can be safely put on a lattice (see Appendix~\ref{appendix:lattice_model} for more details) without acquiring unphysical Dirac points, at $k_{x,y,z} = \pm \pi/a$ for a cubic lattice with lattice constant $a$ for example. The corresponding terms in the Hamiltonian are also known as Wilson mass terms~\cite{Wilson1974, Zhou2017}.
Specific values for the parameters representing various 3D TI materials can be found in Table~\ref{table:params_Zhang}.

When considering a 3D TI slab with surface orthogonal to the $z$~direction with this Hamiltonian, an isotropic gapless surface state spectrum is obtained, described by the following 2D effective Hamiltonian~\cite{shen2012topological}:
%
\begin{equation} \label{eq:Zhang_surface}
	\hat{\mathcal{H}}_\surface^\Zhang = - C_z M_0 / M_z - M_\perp (\hat{p}_x^2 + \hat{p}_y^2) + \sgn(M_z) \sqrt{1 - (C_z/M_z)^2} A_\perp ( \hat{p}_x \sigma_y - \hat{p}_y \sigma_x ).
\end{equation}
\end{widetext}
The wave function profile perpendicular to the $x$-$y$ surface of the $k_x = k_y = 0$ surface state has the following form when the 3D TI is confined to the $z > 0$ region~\cite{Liu2010, shen2012topological}:
\begin{equation} \label{eq:Zhang_SS_solution}
	\begin{split}
		\chi(z) &= ( \, \begin{matrix} c_1 & - c_1 & c_2 & c_2 \end{matrix} \, )^\intercal \, \lef e^{-q_z^+ z} - e^{-q_z^- z} \rig, \\
		q_z^\pm &\equiv \frac{1}{2} \sqrt{\frac{A_z^2}{M_z^2 - C_z^2}} \pm \sqrt{\frac{1}{4}\frac{A_z^2}{M_z^2 - C_z^2} - \frac{M_0}{M_z}},
	\end{split}
\end{equation}
with two independent parameters $c_1$ and $c_2$ (up to normalization). The wave function extends into the bulk with a characteristic penetration depth (or one could say surface state thickness) $\lambda_z$ that can be defined as $\lambda_z \equiv \max\{1/\Re(q_z^+), 1/\Re(q_z^-)\}$. This solution is for confinement along $z$ and analogous solution can be obtained for confinement along $x$ and $y$. The different depth values are presented for the different parameter sets and confinement directions in Table~\ref{table:params_Zhang}. As the band gap of a typical topologically trivial insulator is very large compared to that of the known 3D TIs, hard wall confinement at the 3D TI surfaces, which is understood throughout this text, is an appropriate approximation.

\begin{table*}[htb]
	\begin{tabular}{ l c c c c c }
		\hline \hline
		& Toy model & Bi${}_2$Se${}_3$ (A)~\cite{Zhang2009} & Bi${}_2$Se${}_3$ (B)~\cite{Liu2010} & Bi${}_2$Te${}_3$ & Sb${}_2$Te${}_3$ \\
		\hline
		$M_0$ (eV) & 0.3 & 0.28 & 0.28 & 0.30 & 0.22 \\
		$M_\perp$ (eV$\cdot$\r{A}${}^2$) & 15 & 56.6 & 44.5 & 57.38 & 48.51 \\
		$M_z$ (eV$\cdot$\r{A}${}^2$) & 15 & 10.0 & 6.86 & 2.79 & 19.64 \\
		$A_\perp$ (eV$\cdot$\r{A}) & 3 & 4.1 & 3.33 & 2.87 & 3.40 \\
		$A_z$ (eV$\cdot$\r{A}) & 3 & 2.2 & 2.26 & 0.30 & 0.84 \\
		$C_\perp$ (eV$\cdot$\r{A}${}^2$) & 0 & $-19.6$ & $-30.4$ & $-49.68$ & 10.78 \\
		$C_z$ (eV$\cdot$\r{A}${}^2$) & 0 & $-1.3$ & $-5.74$ & $-6.55$ & 12.39 \\
		$\lambda_z$ (\r{A}) & 10 & 9.01 & 14.09 & / & 36.28 \\
		$\lambda_\perp$ (\r{A}) & 10 & 25.90 & 19.52 & 20.01 & 27.82 \\
		\hline \hline
	\end{tabular}
	\caption{
		The parameters of the effective 3D Hamiltonian for Bi${}_2$Se${}_3$, Bi${}_2$Te${}_3$, and Sb${}_2$Te${}_3$ are listed~\cite{Zhang2009, Liu2010, qi2011topological}, as well as a set of parameters for an isotropic and electron-hole symmetric toy model.
	}
	\label{table:params_Zhang}
\end{table*}

To model 3D TI nanowires with an arbitrary cross section, we consider a tight-binding formulation of the effective 3D Hamiltonian on a cubic lattice with artificial lattice constant $a$~\cite{Konig2008}, with a uniform magnetic field inserted through a standard Peierls substitution.
All (band structure and transport) simulations for this tight-binding model have been carried out with a parallelized implementation of Kwant~\cite{Groth2014, amestoy2001fully, amestoy2006hybrid}, which treats transport with a scattering approach based on a wave-function formulation. The conductance is obtained from the scattering matrix through Landauer's formula. Our approach is therefore limited to elastic scattering, neglecting the impact of, e.g., electron-phonon or Coulomb interactions. Experimentally, this transport behavior should be retrieved when the contacts are close enough to the nanostructure geometry.

\begin{figure}[tb]
	\centering
	\subfigure[\ ]{\includegraphics[width=0.485\linewidth]{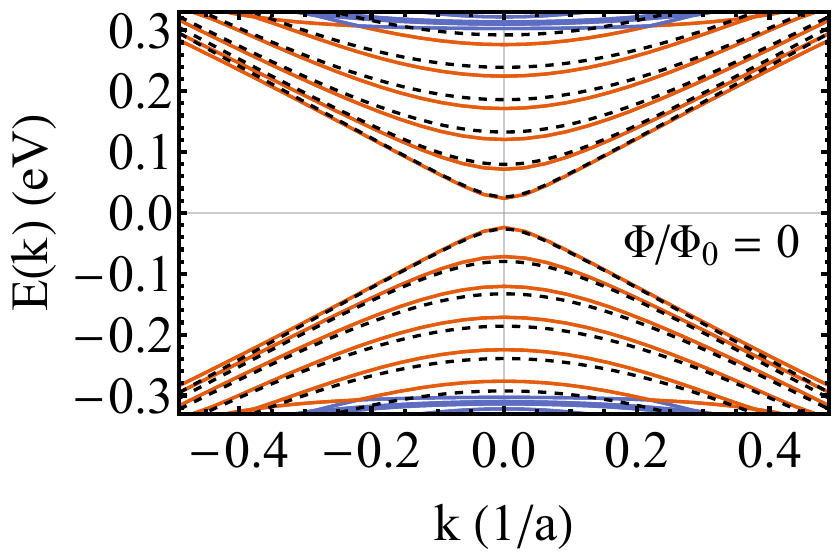}}
	\hspace{0.005\linewidth}
	\subfigure[\ ]{\includegraphics[width=0.485\linewidth]{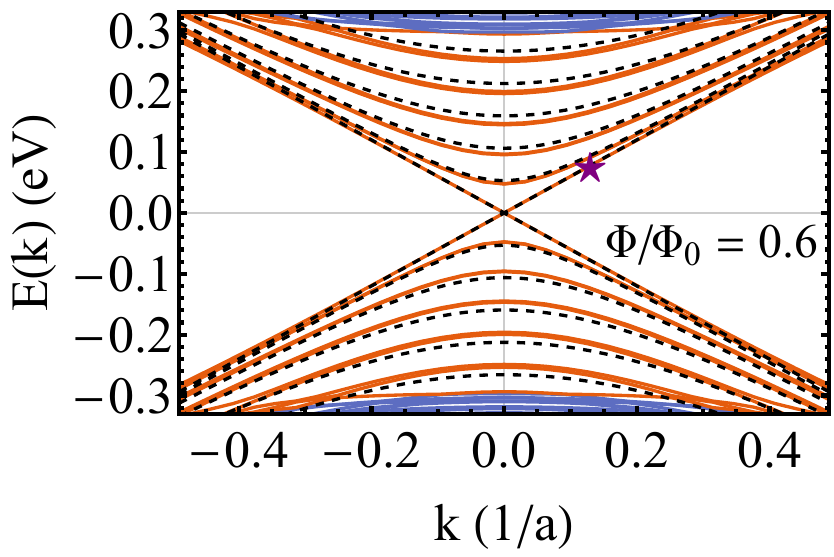}}
	\caption{
		The energy spectrum of a 3D TI nanowire with toy model parameter set of Table~\ref{table:params_Zhang} and a 10$\times$10 nm${}^2$ cross section, considering the effective 3D Hamiltonian with $a = 5$~\r{A}, is shown for (a) zero and (b) \emph{effectively half-integer} ($\Phi/\Phi_0 \approx 0.6$) magnetic flux quantum. The surface (bulk) states are presented in orange (blue). The cylindrical nanowire surface spectrum given by Eq.~(\ref{eq:3D_TI_surface_cyl_en}) is shown in black dashed lines for (a) integer and (b) half-integer generalized angular momenta, considering a cylinder with equal cross section.
		The surface state presented in Fig.~\ref{fig:WF_Cyl_Square}(c) and Fig.~\ref{fig:WF_Cyl_Square}(d) is indicated with a purple star.
	}
	\label{fig:TI_NW_Toy_Spectrum}
\end{figure}

The spectrum of a nanowire with and without magnetic field along the wire axis is presented in Fig.~\ref{fig:TI_NW_Toy_Spectrum}.
On the one hand, the gap in Fig.~\ref{fig:TI_NW_Toy_Spectrum}(a) agrees well with the value of $\hbar \vF / R$ obtained from the 2D Rashba-Dirac model when considering $R = \sqrt{\mathcal{A}/\pi}$ with $\mathcal{A}$ the cross sectional area of the nanowire. This can be expected when the surface states cannot tunnel through the bulk region~\cite{Egger2010, Brey2014}, something which is exponentially suppressed as long as the surface state thickness is significantly smaller than the minimal distance required to cross the bulk region~\cite{Lu2010}.
On the other hand, a minimal total flux of $\Phi/\Phi_0 \approx 0.6$ appears to be required in Fig.~\ref{fig:TI_NW_Toy_Spectrum}(b) to close the gap, deviating slightly from the condition that can be obtained from the Rashba-Dirac spectrum in Eq.~(\ref{eq:3D_TI_surface_cyl_en}). The cause of this offset is found to be the finite thickness of the surface states and will be discussed below.

\begin{figure*}[htb]
	\centering
	\subfigure[\ ]{\includegraphics[height=0.2\linewidth]{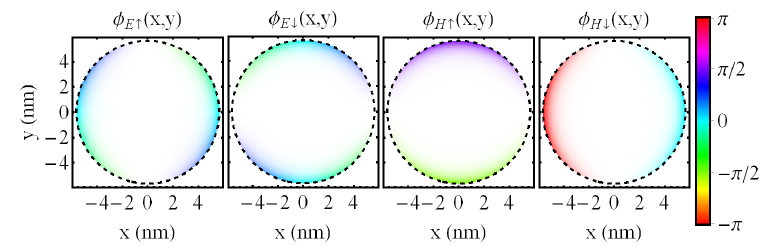}}
	\subfigure[\ ]{\includegraphics[height=0.2\linewidth]{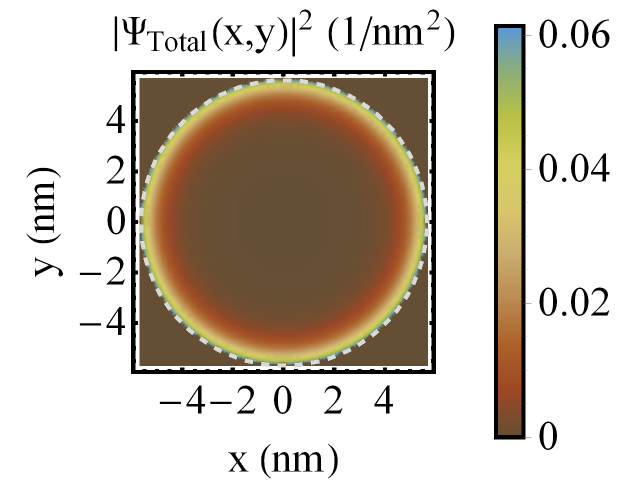}}
	\subfigure[\ ]{\includegraphics[height=0.2\linewidth]{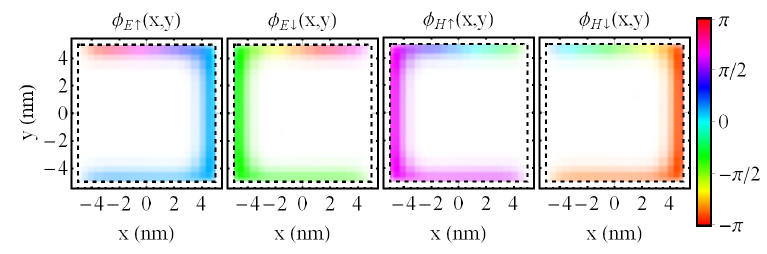}}
	\subfigure[\ ]{\includegraphics[height=0.2\linewidth]{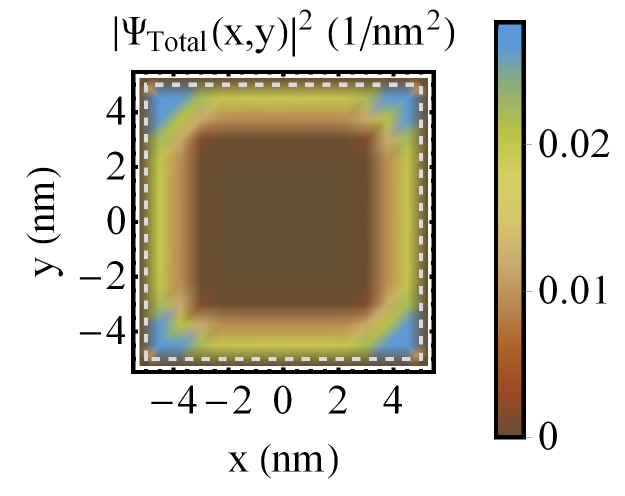}}
	\caption{
		A cross section of the wave function of a gapless subband state [see Fig.~\ref{fig:TI_NW_Toy_Spectrum}(b)] of a (a),~(b) cylindrical (c),~(d) rectangular 3D TI nanowire is presented. (a),~(c) The phase of the different orbitals (spinor components) is shown in color, with the brightness proportional to the local orbital density. (b),~(d) The local density of the total wave function is indicated. The surface of the nanowire is marked with a (a),~(c) black (b),~(d) gray dashed line. (a),~(b) The Rashba-Dirac surface state spinor with $M = 0.3$~eV/$\vF^2$ and $\vF = 3$~eV$\cdot$\r{A} and the (c),~(d) effective 3D Hamiltonian with toy model parameter set of Table~\ref{table:params_Zhang} and $a = 1$~\r{A} are considered, both for a nanowire with cross section equal to $100$~nm${}^2$.
	}
	\label{fig:WF_Cyl_Square}
\end{figure*}

An example of a surface state wave function resulting from the tight-binding model for a 3D TI nanowire with square cross section can be found in Fig.~\ref{fig:WF_Cyl_Square} next to the Rashba-Dirac-based counterpart for a cylindrical nanowire. The local density and phase dependence of the different orbitals are gauge-dependent, but the surface state solutions for orbitals E$\uparrow$ (E$\downarrow$) and H$\downarrow$ (H$\uparrow$) are universally related by a constant phase shift. This relation depends on the geometry of the cross section, as can be understood from Eq.~(\ref{eq:3D_TI_surface_3D}). For a circular cross section, the orbitals differ by a unit of angular momentum provided by $\sigma_r$, while Pauli matrices $\sigma_{x,y}$, carrying zero angular momentum, provide a constant phase shift in case of a square (rectangular) cross section.

For magnetotransport, the surface state thickness will be crucial. It determines the effective piercing magnetic flux, something which cannot be captured by surface models such as the 2D Rashba-Dirac model or other effective 2D models~\cite{Zhang2009, Ostrovsky2010, Liu2010, Brey2014, Zhou2017}. When the nanowire cross section is too small to neglect the surface state thickness, a rescaling of the flux needs to be considered for a precise tuning of the magnitude and orientation of the magnetic field.
A rescaling ratio $\alpha$ can be estimated by $\alpha \approx 1 - \langle \lambda \rangle \mathcal{C} / (2 \mathcal{A}$), with $\langle \lambda \rangle$ the average surface state thickness along the circumference $\mathcal{C}$.
Note that the Rashba-Dirac and 3D model surface states have equal thickness here, but this is generally not the case as their thickness is governed by unrelated parameters in Eqs.~(\ref{eq:3D_TI_surface_3D}) and (\ref{eq:Zhang_SS_solution}), respectively. From Table~\ref{table:params_Zhang}, it is clear that the rescaling ratio $\alpha$ can vary significantly between different 3D TI materials and nanowire orientations. In principle, one should be able to verify this with precise magnetotransport measurements.

\section{Magnetotransport} \label{sec:transport}
We will focus on the magnetotransport properties of three different nanostructures made of connected rectangular nanowires: a straight nanowire, a kink with angle $\gamma_\kink$ (here with $\gamma_\kink = \pi/2$), and a Y-junction (see Fig.~\ref{fig:sketches}). The simulations are limited to nanowires with a uniform square cross section of $10 \times 10$~nm${}^2$ without disorder, considering the toy model and the Bi${}_2$Se${}_3$ (A) parameter set of Table~\ref{table:params_Zhang}.
The main trade-off when considering a larger (smaller) cross sectional area will be a smaller (larger) required magnitude of the magnetic field ($\propto \mathcal{A}$) versus a smaller (larger) energy window ($\propto \sqrt{\mathcal{A}}$) in which the magnetotransport signatures of the gapless subband will appear.
The dependency on cross section size and shape (e.g., the aspect ratio of a rectangular cross section) and disorder has already been investigated and reported in detail elsewhere and will not be discussed further~\cite{Bardarson2010, Iorio2016, Xypakis2017, Ziegler2018}.
For the Bi${}_2$Se${}_3$ transport simulations, the direction of uniaxial anisotropy is considered to be perpendicular to the plane ($x$-$z$) spanned by the legs of the kink or the Y-junction, in line with the experimental feasibility of these structures.
The external magnetic field on the other hand is always considered with in-plane orientation, minimizing its perpendicular component.

\subsection{Straight nanowire} \label{subsec:transport_straight}
\begin{figure}[tb]
	\centering
	\subfigure[\ ]{\includegraphics[width=0.485\linewidth]{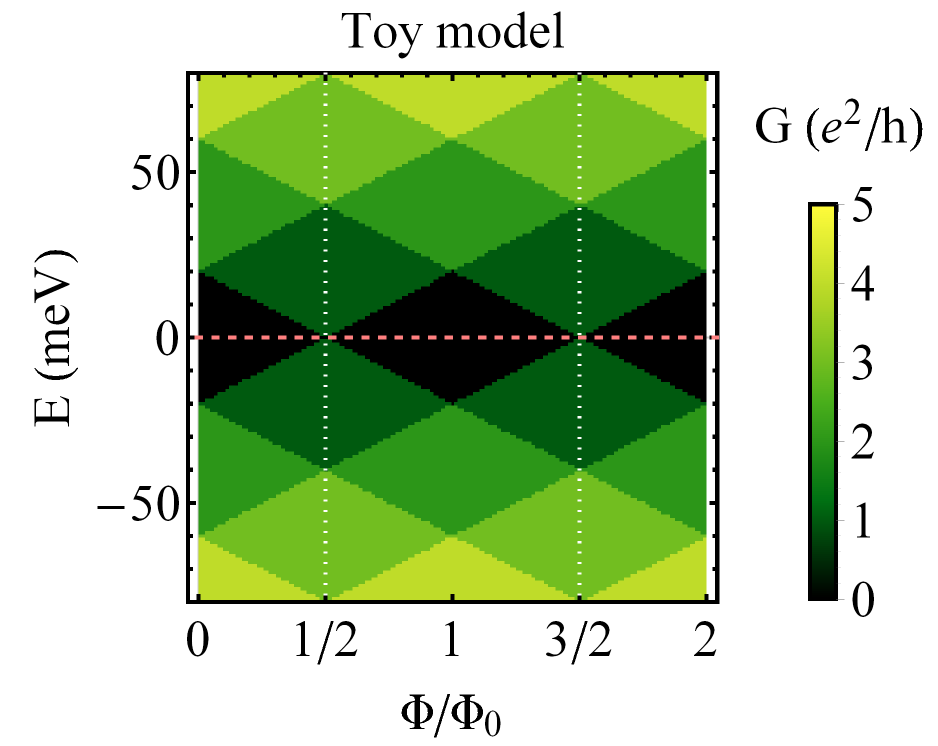}}
	\hspace{0.005\linewidth}
	\subfigure[\ ]{\includegraphics[width=0.485\linewidth]{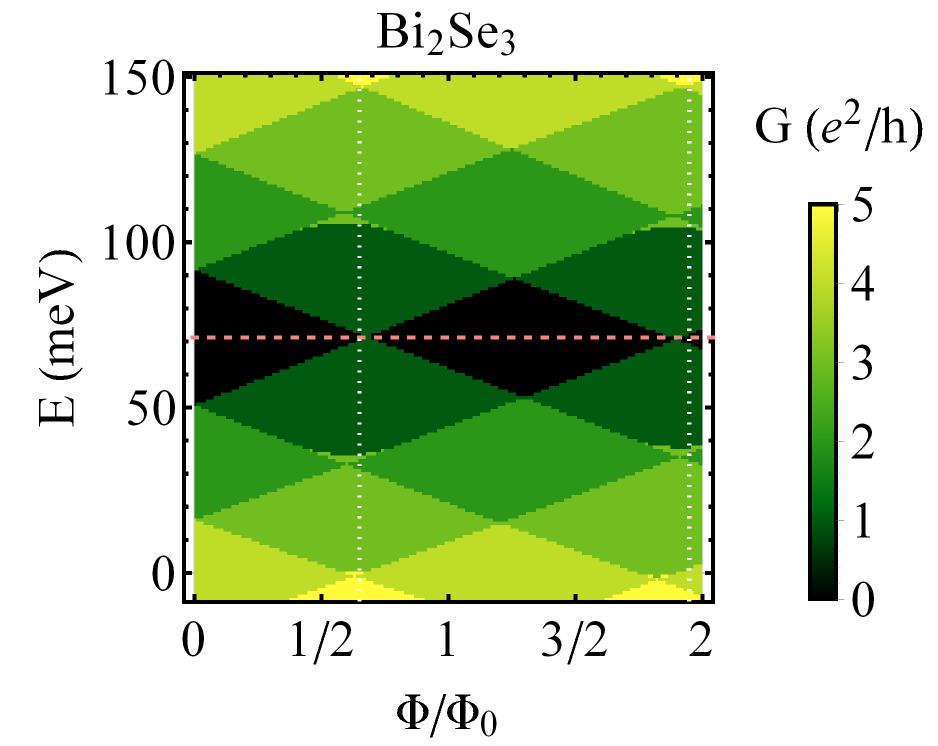}}
	\caption{
		The conductance of a $10\times10$~nm${}^2$ 3D TI nanowire is shown as a function of the energy (near the Dirac point) and magnetic flux $\Phi$ from a fully aligned uniform magnetic field. The results were obtained with a tight-binding version of the effective 3D Hamiltonian presented in Eq.~(\ref{eq:ham_Zhang}) with the (a) toy model (b) Bi${}_2$Se${}_3$ (A) parameter set of table~\ref{table:params_Zhang} and $a=10$~\r{A}.
		The Dirac point energy equal to (a) 0~meV (b) 71~meV and the effective half-integer magnetic fluxes with rescaling ratio (a) $\alpha = 1$ (b) $\alpha = 0.77$ are indicated with pink dashed lines and white dotted lines, respectively.
	}
	\label{fig:conductance}
\end{figure}

The conductance of a straight 3D TI nanowire with toy model and Bi${}_2$Se${}_3$ parameter sets is presented in Fig.~\ref{fig:conductance}. The Dirac point is centered at 0~meV for the electron-hole symmetric toy model and near 71~meV for Bi${}_2$Se${}_3$. The value of the latter is well estimated by $-M_0(C_\perp/M_\perp + C_z/M_z)/2 \approx 67$~meV, being the average of the Dirac point energy for TI slab surface states parallel to the $x$-$z$ and $y$-$z$ planes [see Eq.~(\ref{eq:Zhang_surface})], respectively.
The typical diamond tile pattern for the magnetoconductance is clearly visible in both cases and the electron-hole asymmetry of Bi${}_2$Se${}_3$ is barely visible. The flux rescaling ratio $\alpha$, which can be extracted from the conductance profile, is significantly smaller for the Bi${}_2$Se${}_3$ parameter set, as expected from the estimate $\alpha \approx 1 - \langle \lambda \rangle \mathcal{C} / (2 \mathcal{A})$ because $\langle \lambda \rangle$ is larger for Bi${}_2$Se${}_3$. The agreement between this estimate for $\alpha$ and its fitted value from the conductance profile is not perfect, however, because the lattice constant in our simulations, $a = 10$~\r{A}, is too large for an accurate retrieval of the surface state depth profile.

\subsection{Kink} \label{subsec:transport_kink}
Compared to straight wires, a much richer magnetotransport behavior can be expected for kinks. A priori, the lack of translational invariance and time-reversal symmetry allows for elastic backscattering of any surface state, even in ideal nanowire kinks without disorder. It has already been shown theoretically that, for 3D TI slabs which are tilted with respect to each other, angle-dependent reflections will occur at their interface~\cite{Sen2012}. In this work, we focus on the nanostructure regime where well-separated subbands due to confinement play an important role. As is the case for straight TI nanowires, the confinement gap can be closed in both legs of the kink simultaneously by applying an external magnetic field with the appropriate magnitude under the correct angle $\gamma_B$. This can be translated to the following condition:
\begin{equation} \label{eq:zero_gap_cond}
	\frac{\cos\gamma_B}{2 n_\mathrm{I} + 1} = \frac{\cos(\gamma_\kink - \gamma_B)}{2n_\mathrm{O} + 1},
\end{equation}
with $n_{\mathrm{I}, \mathrm{O}}$ integers such that $\Phi_{\mathrm{I}, \mathrm{O}} = (2 n_{\mathrm{I}, \mathrm{O}} + 1) \Phi_0 / 2$,
where $\Phi_\mathrm{I} = |\vecB \cos\gamma_B| \mathcal{A}$ and $\Phi_\mathrm{O} = |\vecB \cos(\gamma_\kink - \gamma_B)| \mathcal{A}$ are the piercing magnetic fluxes of the input and output leg, respectively (see Fig.~\ref{fig:sketches}).
Evidently, unlike for straight TI nanowires, the appearance of a perpendicular component of a uniform external magnetic field cannot be prevented throughout the whole structure, but the gapless Dirac spectrum is expected to survive as long as the magnitude of the perpendicular component stays below $B_\crit \equiv \pi \Phi_0 / (2 \mathcal{A})$, in analogy to the perturbative result based on the 2D Rashba-Dirac model for a cylindrical nanowire presented in Sec.~\ref{subsec:NW_cyl}.

\begin{figure}[tb]
	\centering
	\subfigure[\ ]{\includegraphics[width=0.485\linewidth]{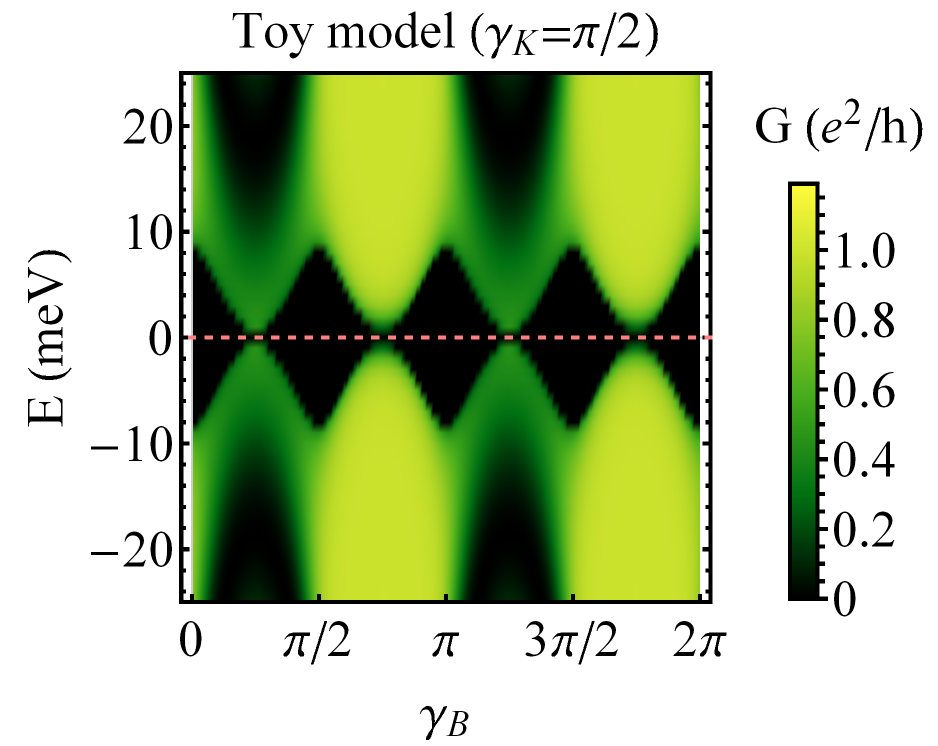}}
	\hspace{0.005\linewidth}
	\subfigure[\ ]{\includegraphics[width=0.485\linewidth]{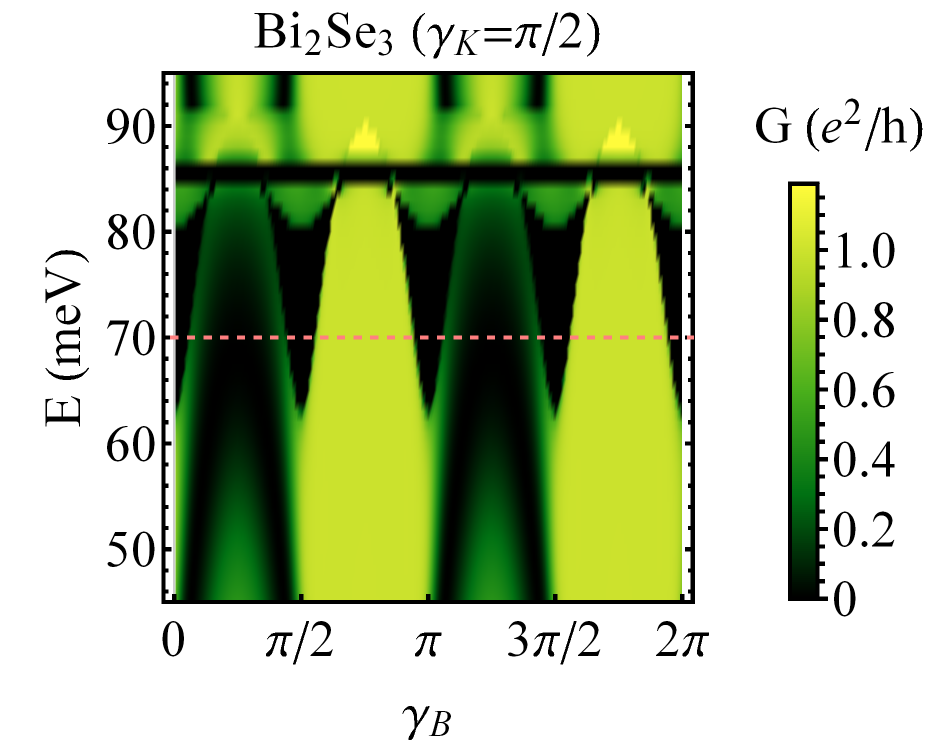}}
	\caption{
		The conductance of a 3D TI nanowire kink, with $\gamma_\kink = \pi/2$ and both legs effectively pierced by a half-integer magnetic flux quantum, is shown as a function of the angle $\gamma_B$ of the applied magnetic field ($|\vecB| = \sqrt{2} \Phi_0 / (2 \alpha \mathcal{A})$) and the (equal) energy level of the leads. The nanowire parameters for (a) and (b) are the same as in Fig.~\ref{fig:conductance} and the same Dirac point energy is indicated with a pink dashed line.
	}
	\label{fig:90_deg_kink_conductance}
\end{figure}

We proceed by considering a kink with fixed angle $\gamma_\kink$ while allowing the magnetic field to rotate, the setup which is most easily set up experimentally. When $\gamma_\kink = \pi/2$ and the magnitude of the magnetic field is tuned to $\sqrt{2} \Phi_0 / (2 \alpha \mathcal{A})$, the perpendicular component is within limits (when $\alpha$ is reasonably close to 1) and a gapless subband channel should appear in both legs of the kink when $\gamma_B = \pi/4 + m \pi/2$ with $m \in \mathbb{Z}$. Hence, for these angles, one could expect transmission across the kink at energies arbitrarily close to the Dirac point.
Results of conductance simulations for this system are presented in Fig.~\ref{fig:90_deg_kink_conductance} and the expected transmission behavior can indeed be identified.

Interestingly, in an energy window around the Dirac point that is of the same order as the confinement gap, the transmission probability appears to be very weak for $m$ even and perfect for $m$ odd.
This implies a strong dependence on the relative orientation of the magnetic field with respect to the input and output legs, for which the following general behavior can be identified.
We can distinguish two types of relative orientations of the magnetic field: an \emph{aligned} (for the $\gamma_\kink = \pi/2$ kink, when $0 < \gamma_B < \pi/2$ or $\pi < \gamma_B < 3\pi/2$) and a \emph{transverse} orientation (for the $\gamma_\kink = \pi/2$ kink, when $\pi/2 < \gamma_B < \pi$ or $3\pi/2 < \gamma_B < 2\pi$), where aligned (transverse) refers to the orientation of the magnetic field with respect to the transport direction from input to output.
Gap-closing conditions at the input and output with transverse ($m$ odd) orientation appear to instigate maximal overlap (perfect transmission) between the input and output states of the gapless helical subband, while the aligned orientation instigates minimal overlap (close to zero transmission).
This behavior appears to hold for kinks with arbitrary angles, both for the toy model and the Bi${}_2$Se${}_3$ parameter set, as long as the perpendicular component of the magnetic field remains in the perturbative regime. Note that a straight wire cannot have gap-closing conditions with a transverse magnetic field orientation.

For the Bi${}_2$Se${}_3$ kink, the electron-hole asymmetry becomes noticeable in the conductance near the Dirac point, unlike for a straight nanowire. The gap closes asymmetrically and the closing point is shifted about 15~meV in energy as compared to the Dirac point of the straight wire with an aligned magnetic field. A gap persists at the gap-closing angles, implying that the symmetric Dirac velocity renormalization, as derived perturbatively above for perpendicular magnetic fields, is not valid for an electron-hole asymmetric and/or anisotropic 3D TI Hamiltonian. However, this gap is only a couple of meV, being much smaller than the confinement gap, and does not prevent a clear perfect transmission signature from showing close to (mostly below) the (shifted) Dirac point in case of a parallel magnetic field orientation.

\subsection{Y-junction} \label{subsec:transport_junction}
In this section, a 3D TI Y-junction geometry [see Fig.~\ref{fig:sketches}(b)] is considered, with three nanowire legs having identical cross sections. Led by the magnetoconductance behavior for kinks in the previous section, we can already expect that a properly tuned and oriented external magnetic field should also be able to realize (nearly) gapless perfect transmission between any selection of the three legs of the Y-junction. A T-junction, for example, would not offer the same flexibility, as two of the three legs always align identically with the applied magnetic field, as is the case for a straight nanowire.

\begin{figure*}[htb]
	\centering
	\subfigure[\ ]{\includegraphics[height=0.2\linewidth]{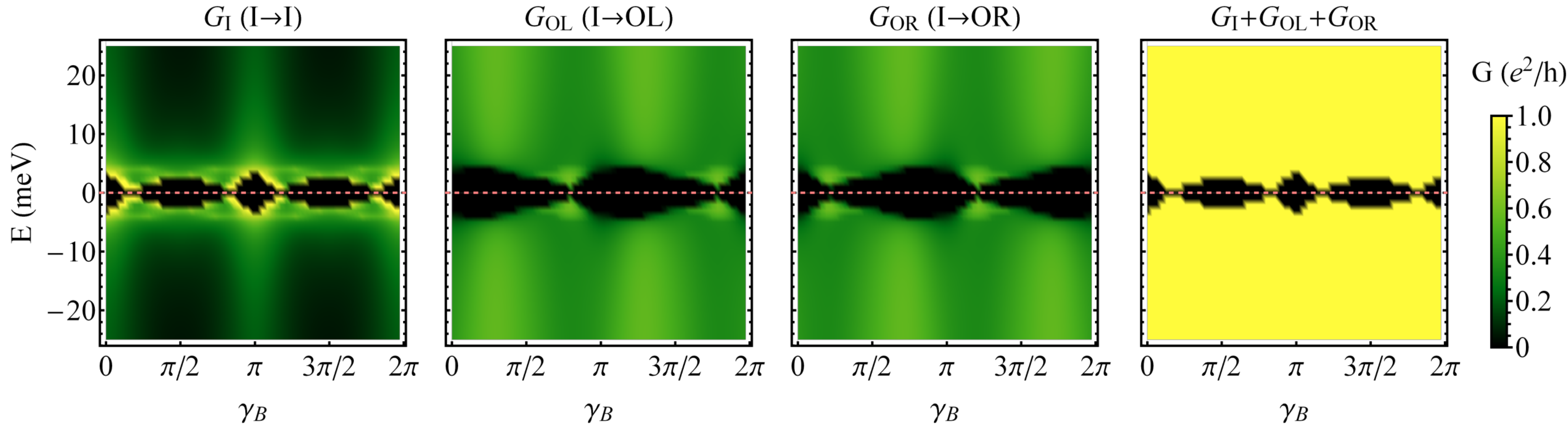}}
	\hspace{0.0\linewidth}
	\subfigure[\ ]{\includegraphics[height=0.2\linewidth]{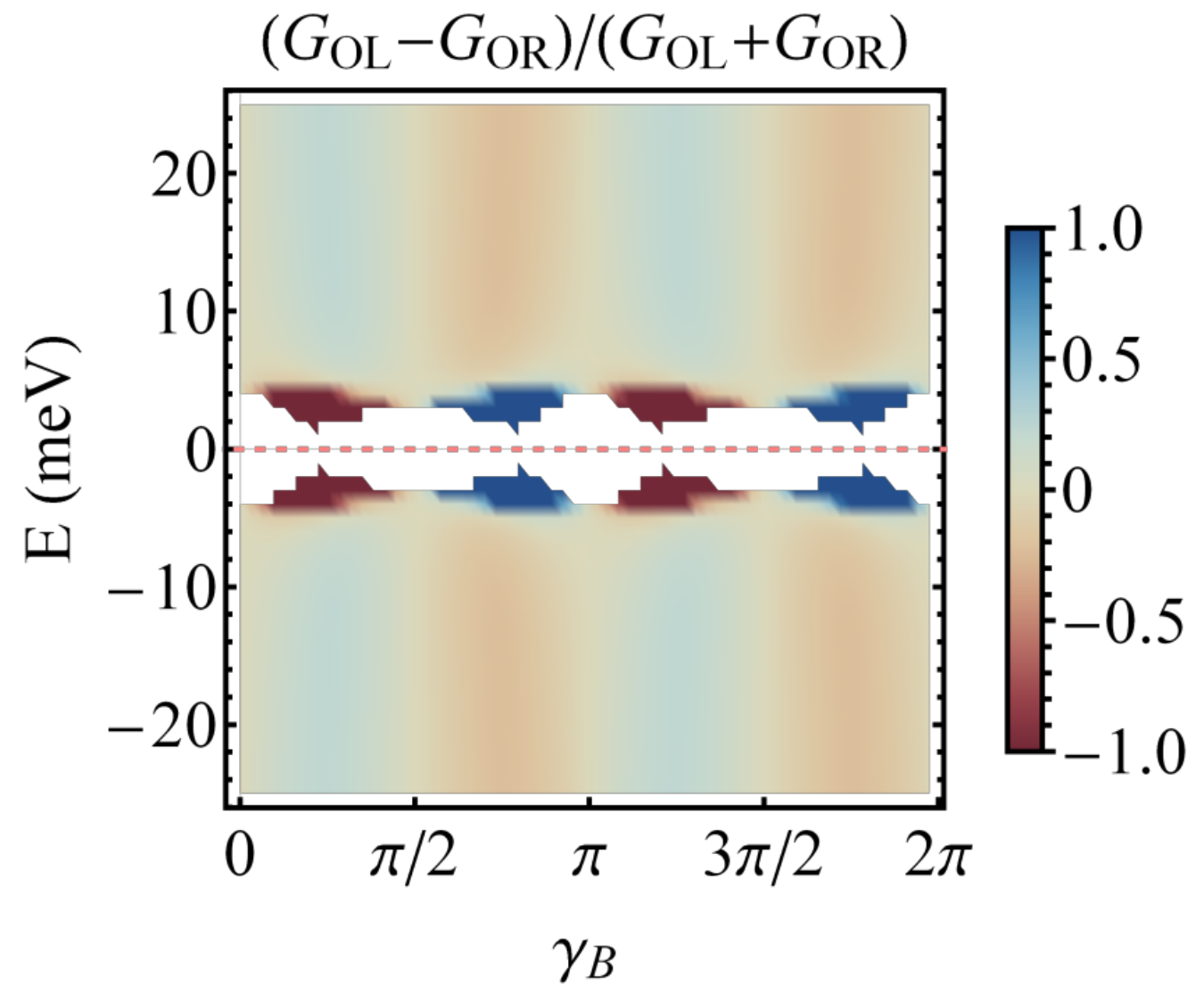}}
	\subfigure[\ ]{\includegraphics[height=0.2\linewidth]{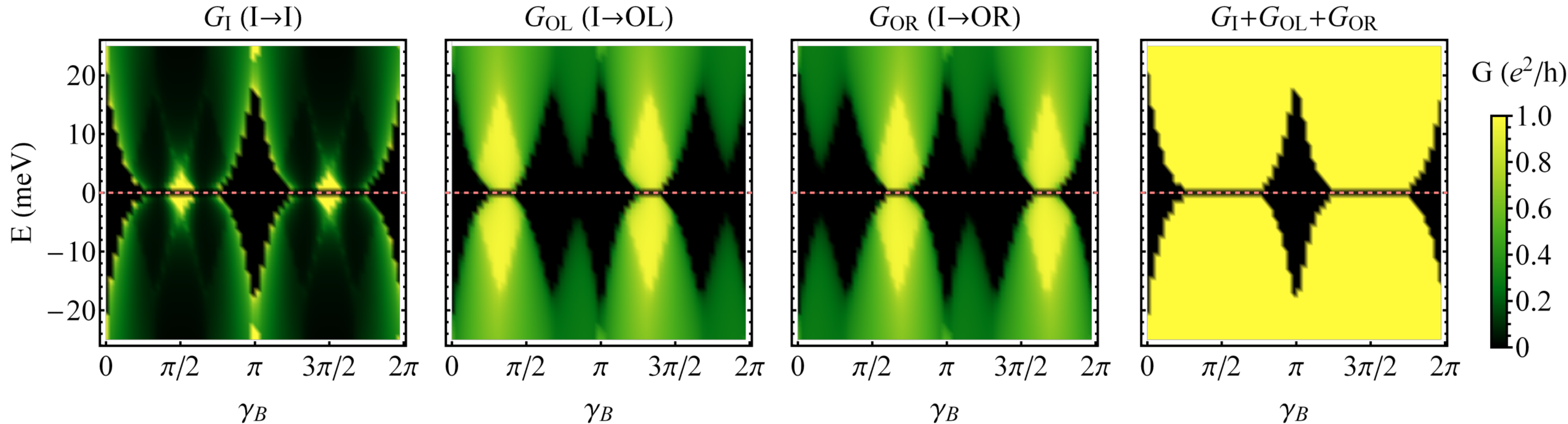}}
	\hspace{0.00\linewidth}
	\subfigure[\ ]{\includegraphics[height=0.2\linewidth]{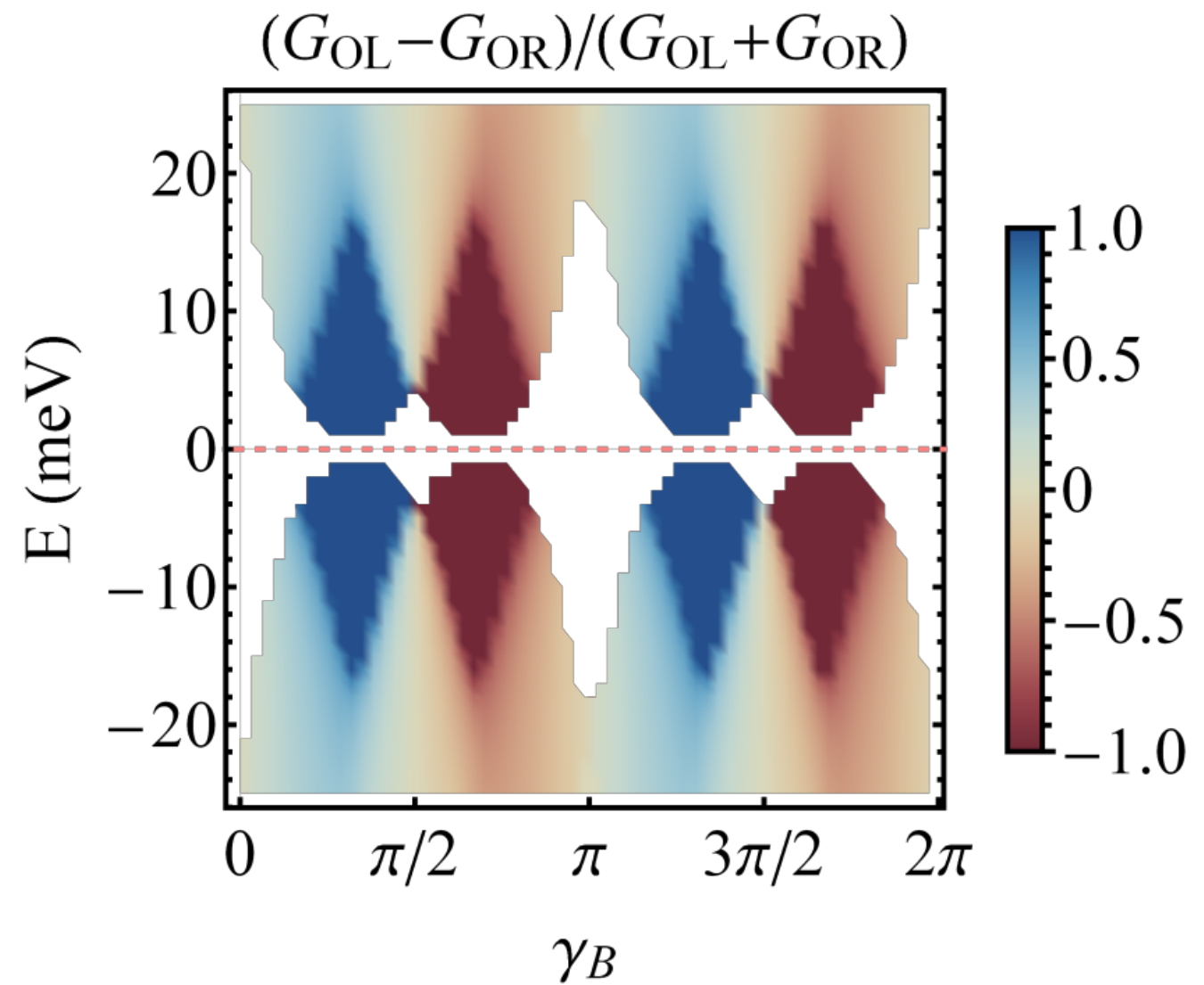}}
	\subfigure[\ ]{\includegraphics[height=0.2\linewidth]{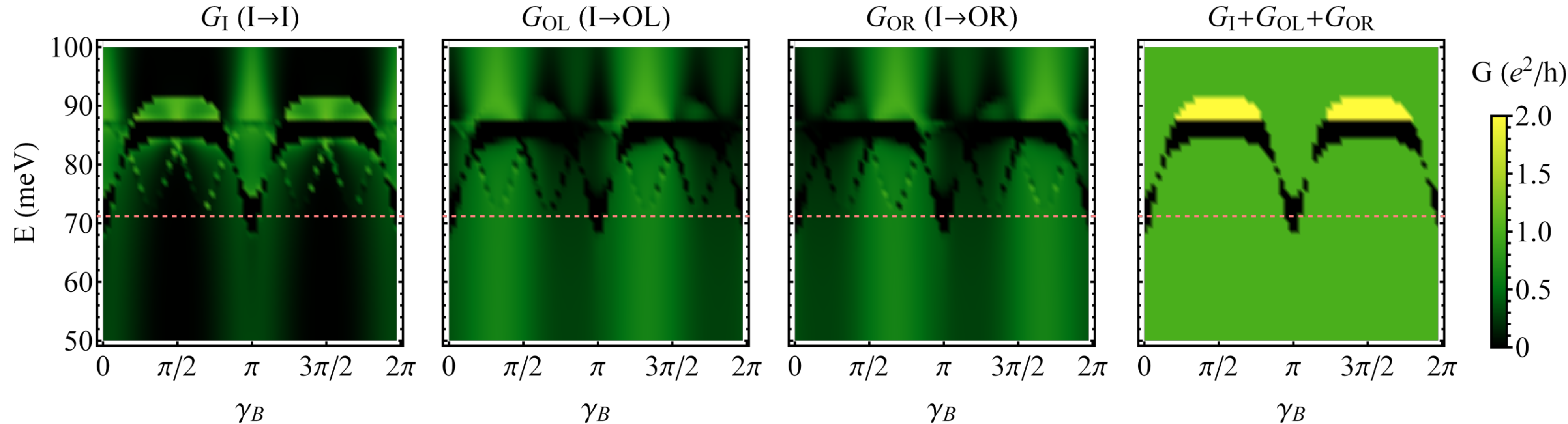}}
	\hspace{0.00\linewidth}
	\subfigure[\ ]{\includegraphics[height=0.2\linewidth]{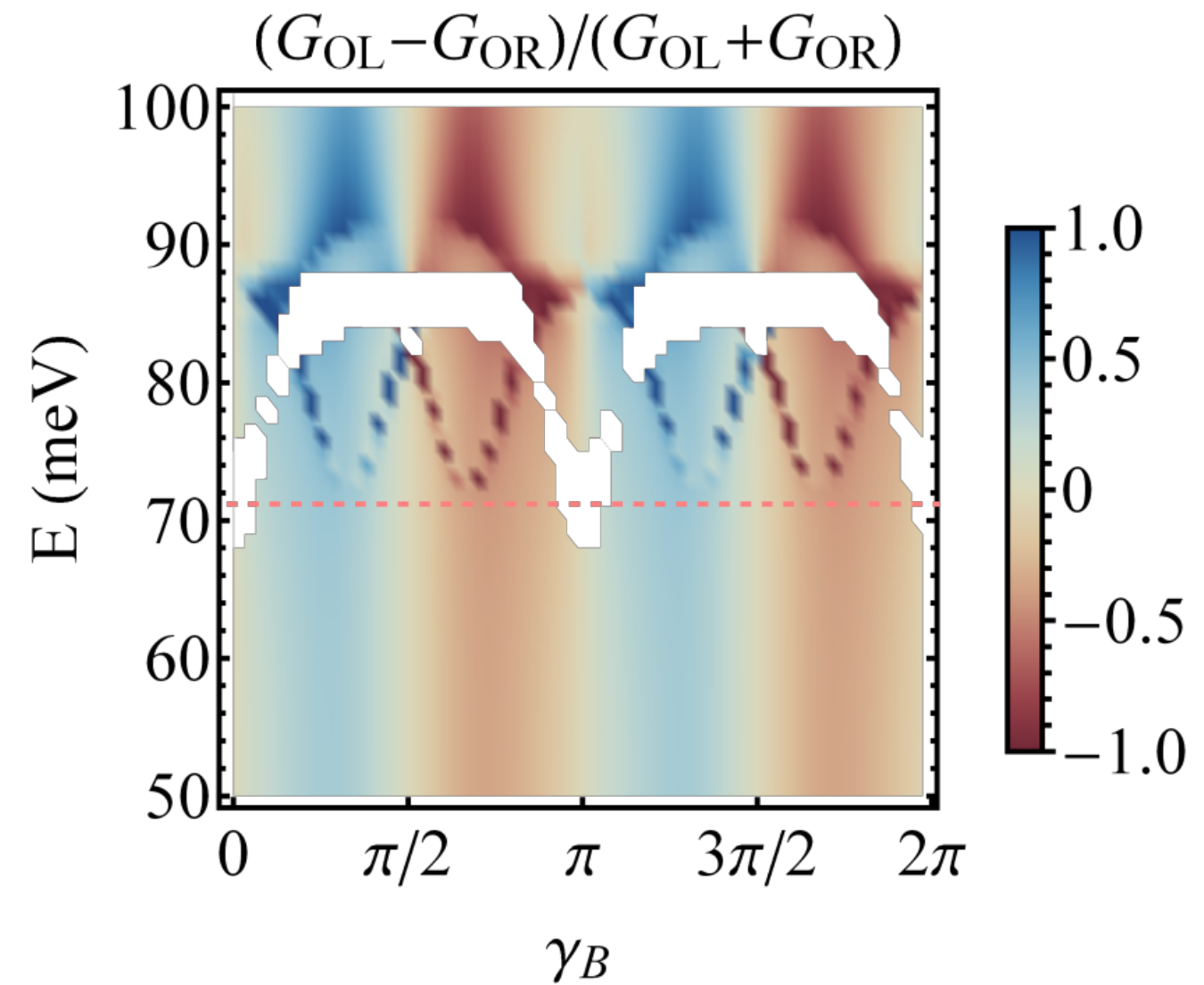}}
	\subfigure[\ ]{\includegraphics[height=0.2\linewidth]{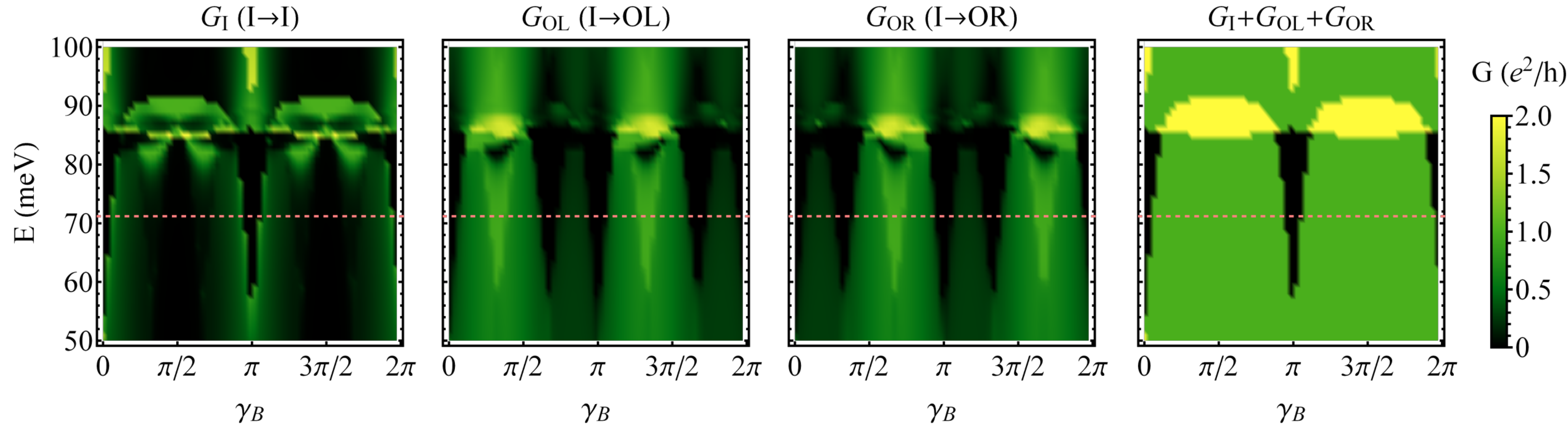}}
	\hspace{0.00\linewidth}
	\subfigure[\ ]{\includegraphics[height=0.2\linewidth]{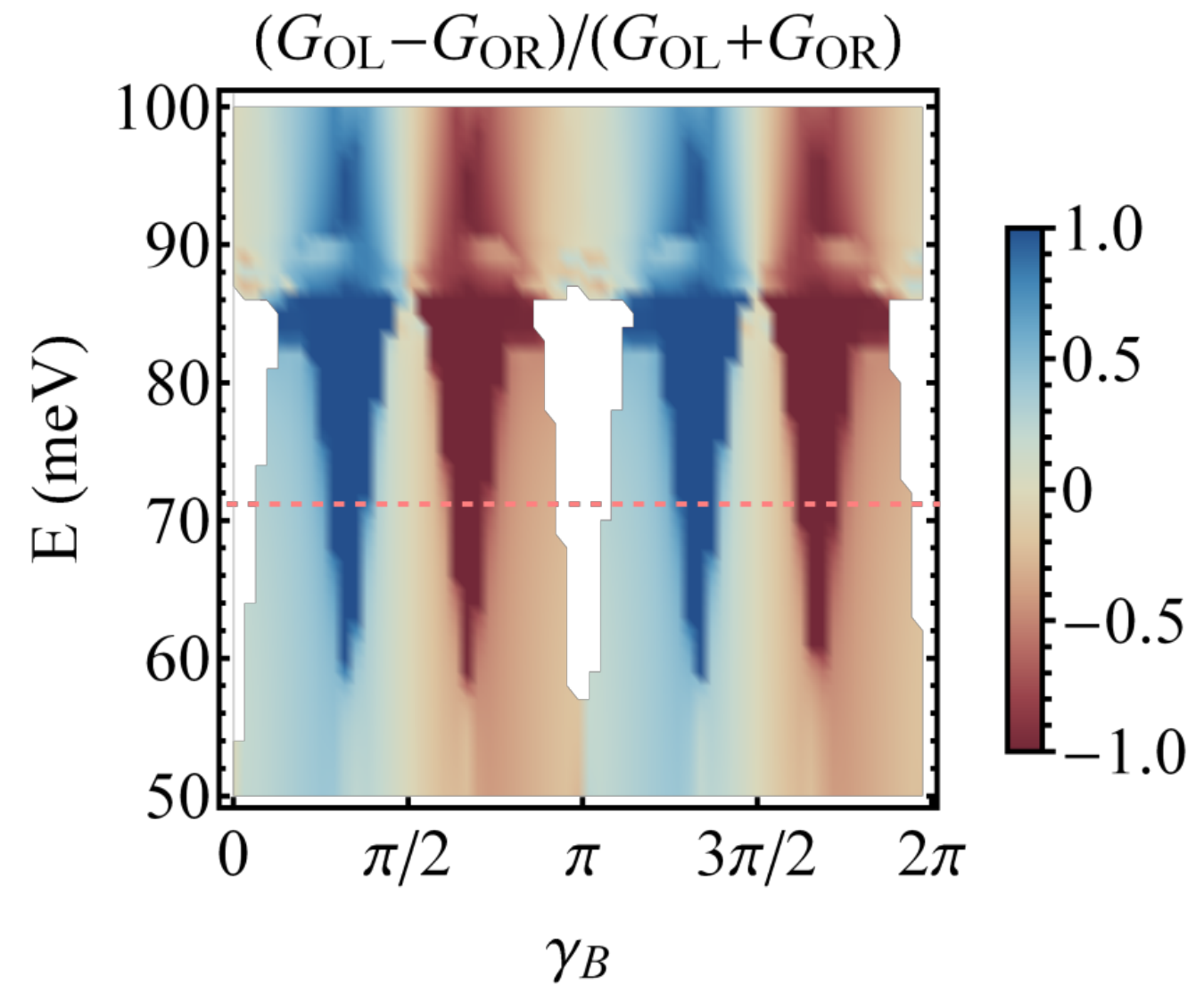}}
	\caption{
		(a),~(c),~(e),~(g) The conductance between the different legs of a 3D TI Y-junction and (b),~(d),~(f),~(h) the splitting ratio for the current between the left and right output legs are shown as a function of energy and the angle $\gamma_B$ of the magnetic field, with its magnitude equal to (a)-(b),~(e)-(f) $|\vecB| = \Phi_0 / (\sqrt{3} \alpha \mathcal{A})$ (c)-(d),~(g)-(h) $|\vecB| = \Phi_0 / (\alpha \mathcal{A})$. The (a)-(d) toy model (e)-(h) Bi${}_2$Se${}_3$ (A) parameter set is considered, with the remaining parameters identical to those considered in Fig.~\ref{fig:conductance}. The Dirac point energy of the corresponding straight TI nanowire is indicated with a pink dashed line. The angle of $\pi/3$ between the $z$~axis and the two output channels was approximated by $\arctan(3/5)$ in the tight-binding simulations for sufficient lattice commensurability.
	}
	\label{fig:Y_junction_conductance}
\end{figure*}

The results for conductance simulations of a Y-junction are shown in Fig.~\ref{fig:Y_junction_conductance}, with an external magnetic field having a magnitude of $\Phi_0 / (\alpha \mathcal{A})$ or $\Phi_0 / (\sqrt{3} \alpha \mathcal{A})$. These magnitudes lead to simultaneous gap-closing conditions in two of the three legs with \emph{transverse} or \emph{aligned} orientation of the magnetic field (the third leg lying parallel or perpendicular to it), respectively.
Note that both orientations require a different magnitude of the magnetic field, unlike for the right-angle kink. Both magnitudes are within limits for the perpendicular component, however, such that the phenomenology of the gapless helical subband should survive.
From the transport simulation results, different conductance regimes can be clearly identified near the Dirac point for both the toy model and the Bi${}_2$Se${}_3$ Y-junction: a fully gapped regime, a reflection-dominated regime, and left- and/or right-transmitting regimes.
The magnitude of $\Phi_0 / (\sqrt{3} \alpha \mathcal{A})$ leads to gap-closing conditions in two of the three legs. These conditions coincide with an \emph{aligned} orientation of the magnetic field. Unlike for the kink, however, this does not lead to a much reduced transmission when compared to the \emph{transverse} orientation. The transverse orientation exhibits a strong directionality of the current, however. The transmitted current can be perfectly steered to one of the two output legs, as can be seen in the splitting ratio.

\begin{figure}[tb]
	\centering
	\subfigure[\ ]{\includegraphics[width=0.485\linewidth]{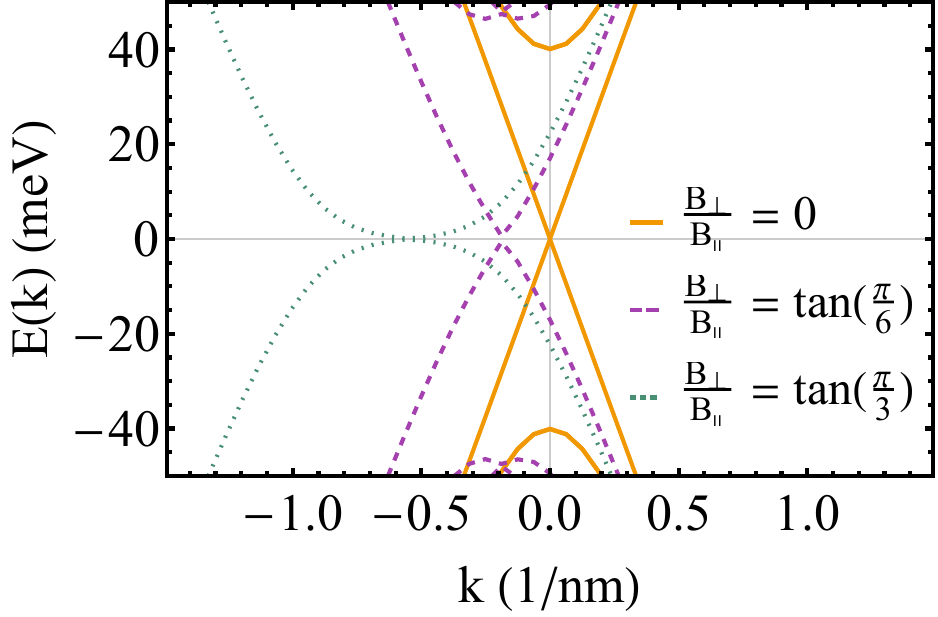}}
	\hspace{0.005\linewidth}
	\subfigure[\ ]{\includegraphics[width=0.485\linewidth]{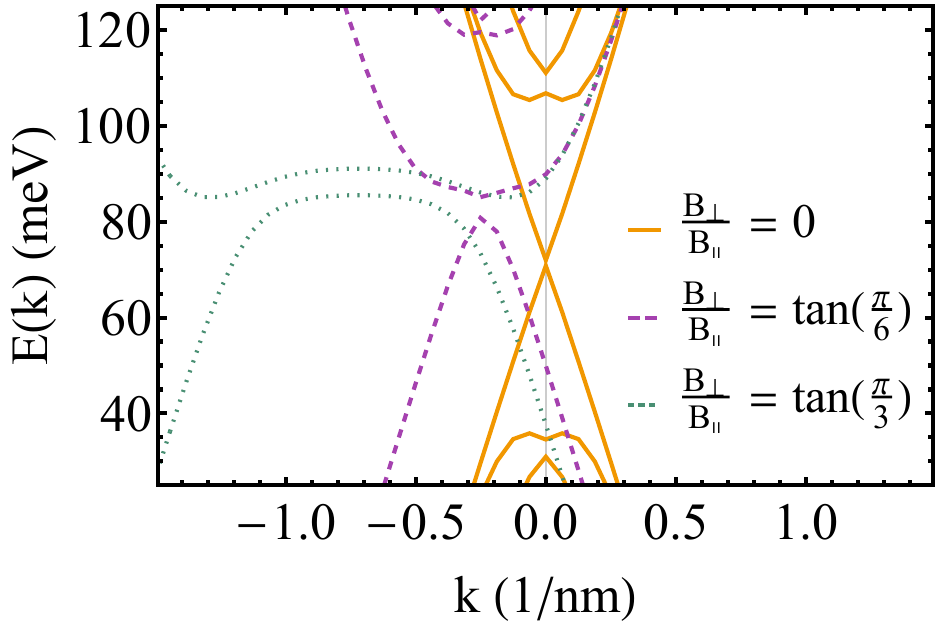}}
	\caption{
		The surface state spectrum of a half-integer flux quantum-pierced 3D TI nanowire with different values for the perpendicular component of the magnetic field, considering (a) the isotropic and electron-hole symmetric toy model and (b) the anisotropic and electron-hole asymmetric Bi${}_2$Se${}_3$ (A) parameter set of Table~\ref{table:params_Zhang} and a 10$\times$10 nm${}^2$ cross section. (b) A small gap opens up and a Mexican-hat-shaped subband develops when the perpendicular component increases.
	}
	\label{fig:Mexican_hat}
\end{figure}

Similar to the results of the right-angle kink, the main difference between the toy model and the Bi${}_2$Se${}_3$ simulations is the particle-hole asymmetry in the conductance signature and an upward shift in energy of about 15~meV for the gap closing with respect to the straight nanowire Dirac point energy. In addition, the perpendicular component of the magnetic field induces a Mexican-hat shape for the subband just above the Dirac point (see Fig.~\ref{fig:Mexican_hat}), leading to a small energy window in which the total conductance is doubled. For the transverse orientation, the top of the lower subband shifts above the bottom of the Mexican-hat-shaped upper subband, such that the minigap, otherwise appearing for Bi${}_2$Se${}_3$ at the gap-closing condition, disappears.

\section{Conclusions and outlook} \label{sec:conclusion}
Based on an analysis of the 2D surface Rashba-Dirac model and tight-binding simulations of an effective 3D model, we have studied the magnetotransport properties of patterned 3D TI nanostructures; in particular, a right-angle kink and a Y-junction made of rectangular nanowires. A perturbative treatment for the 2D Rashba-Dirac model shows that the magnetic flux-driven gap closings and their resulting conductance signatures survive as long as the perpendicular component of the magnetic field, which is unavoidable for this type of structure, stays below a critical value. This result is confirmed with an effective 3D model, which also accounts for the impact of the surface state thickness, the cross sectional shape of the nanowire, and anisotropy, and/or electron-hole asymmetry of the band structure. Compared to effective surface models, the surface state thickness induces a rescaling of the effective piercing magnetic flux, which in turn governs the magnetic field that is required to induce gapless helical subbands.

We demonstrated that, while backscattering is, in principle, allowed in the presence of a perpendicular magnetic field component, perfect (nearly) gapless transmission can be realized near the Dirac point between a certain input and output leg of a kink by applying an external magnetic field with appropriately tuned magnitude and orientation. Apart from piercing the input and output channels with a half-integer magnetic flux quantum to close the confinement gap, an appropriate alignment of the magnetic field is required for maximal overlap of the input and output states and perfect transmission. This is realized by a magnetic field with a \emph{transverse} (rather than an \emph{aligned}) orientation with respect to the transport direction, something which is impossible to realize with straight nanowires.

\begin{figure}[tb]
	\centering
	\subfigure[\ ]{\includegraphics[width=0.3\linewidth]{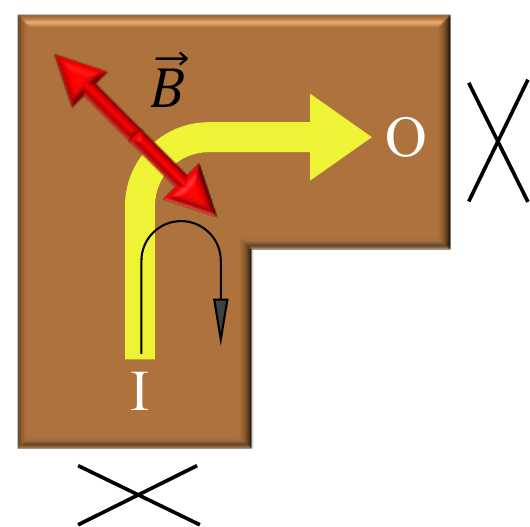}}
	\hspace{0.15\linewidth}
	\subfigure[\ ]{\includegraphics[width=0.3\linewidth]{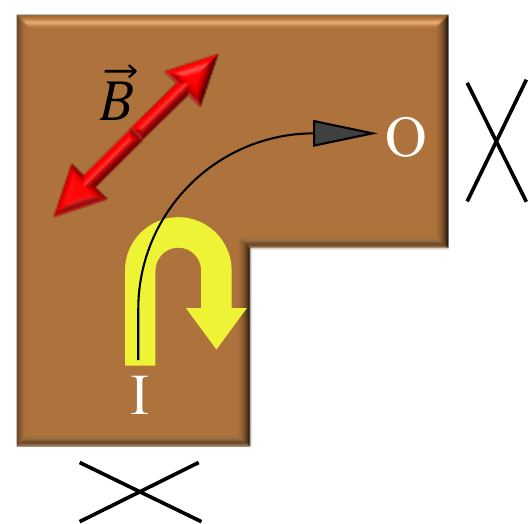}}
	\subfigure[\ ]{\includegraphics[width=0.4\linewidth]{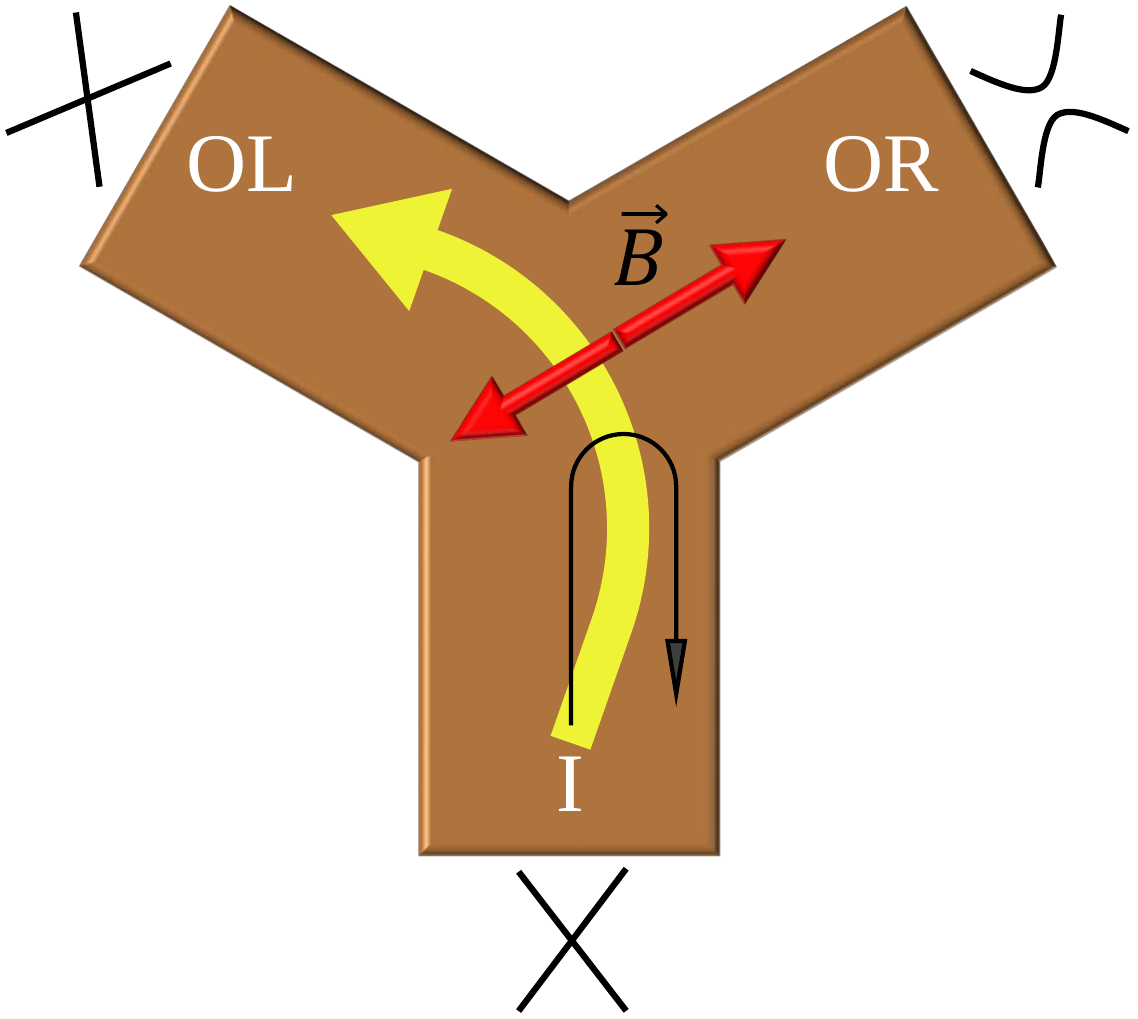}}
	\hspace{0.05\linewidth}
	\subfigure[\ ]{\includegraphics[width=0.4\linewidth]{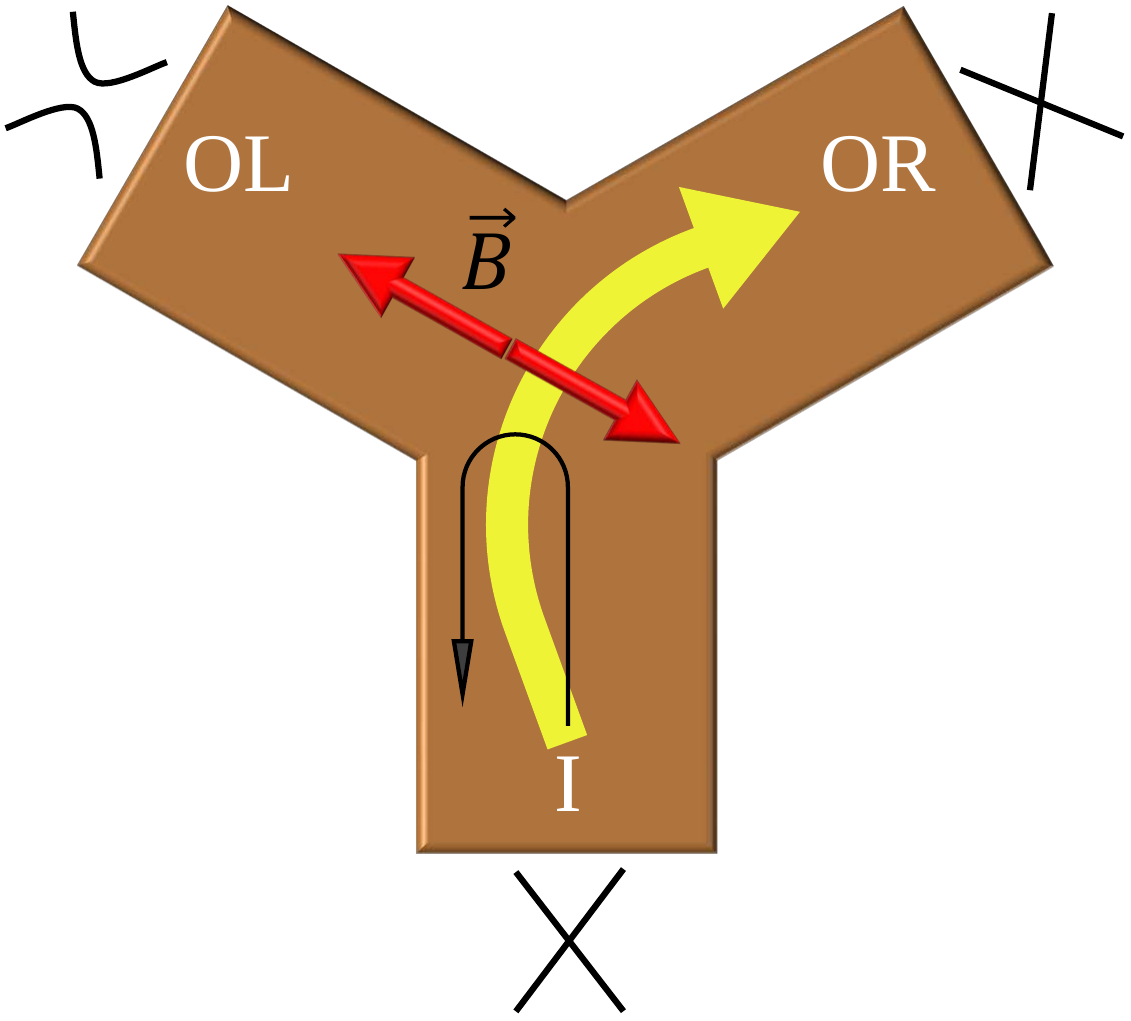}}
	\caption{
		(a),~(b) A right-angle kink with gap-closing conditions in the input and output legs is depicted with (a) a \emph{transverse} and (b) an \emph{aligned} orientation of the external magnetic field, always making an angle of 45 degrees with the transport direction. In case of a(n) transverse (aligned) orientation, the transmission of the gapless helical subband near the Dirac point is perfect (almost zero). As the magnitude of the parallel and perpendicular magnetic field components is the same for both orientations, a difference in magnetoconductance can only be attributed to the topological spin-momentum locked surface states.
		(c),~(d) A Y-junction with gap-closing conditions in the input and (c) left or (d) right output leg is depicted with a transverse orientation of the external magnetic field. This leads to a strong gapless transmission signature near the Dirac point, which can be perfectly steered to one of the two output legs.
	}
	\label{fig:90_deg_kink}
\end{figure}

The difference in magnetoconductance between an aligned and transverse orientation of the magnetic field depends crucially on the spin-momentum locking properties (helicity) of the topological surface states.
Spin-momentum locking implies that a certain change of the direction of momentum should be accompanied by the same change of the direction of spin, something which is optimally furnished by an aligned orientation of the external magnetic field. For trivial surface or bulk states, there is no such requirement and a dependence on the alignment of the magnetic field is therefore absent.
Hence, comparing the magnetoconductance of a 3D TI nanostructure with aligned versus transverse orientation of the external magnetic field offers a new direct experimental probe to identify and characterize magnetotransport of these topological surface states, while the transport is polluted by trivial surface or bulk states.
For a right-angle kink, the only change that is required in the system to compare aligned and transverse orientations of the magnetic field at the gap-closing condition is a 90-degree rotation of the sample with respect to the external magnetic field (see Fig.~\ref{fig:90_deg_kink}). As a function of the in-plane angle of the magnetic field, the simulation results show an indicative $\pi$-periodic magnetoconductance signature, rather than a trivial $\pi/2$-periodic profile.

For a Y-junction, we demonstrated that the transport near the Dirac point can be perfectly steered to either of the two output legs when the magnetic field realizes gap-closing conditions in two of the three legs with a transverse orientation of the magnetic field (see Fig.~\ref{fig:90_deg_kink}). A comparison between the aligned and transverse orientation is less straightforward, as the magnitude of the magnetic field also needs to be changed. This could also affect the behavior of the trivial bulk and surface states, which might hamper the extraction of the topological surface state current from experimental conductance measurements.

3D TI nanostructures such as kinks and Y-junctions provide a new way to explore the (magneto)transport properties of 3D TI surface states and we have presented a fabrication method in detail to realize these structures experimentally and perform quantum transport measurements. This could also lead to new possibilities for applications. The 3D TI Y-junction, for example, has already been proposed by Cook \emph{et al}.\ in combination with a rotating magnetic field and proximity-induced $s$-wave superconductivity to move Majorana bound states between the different ends of the legs~\cite{Cook2012}.
It would be interesting to investigate whether these magnetotransport signatures could be exploited for electrical detection of Majorana bound states in Y- or T-junction~\cite{Plissard2013, Weithofer2014} configurations with proximity-induced superconducting regions, which can be scaled up to networks in a straightforward manner, allowing for fault-tolerant computation schemes~\cite{Alicea2011, Hyart2013, Aasen2016, Litinski2017}. These considerations and an analysis of the robustness of the results against disorder will be investigated in future work.

\appendix

\section{Lattice model}
\label{appendix:lattice_model}
The Hamiltonian of Eq.~(\ref{eq:ham_Zhang}) can be put on a four-orbital square lattice with lattice constant $a$ (analogously to the procedure by Rod \emph{et al}.\ for the two-orbital BHZ Hamiltonian for example~\cite{Bernevig2006,Rod2015}) by making the following substitutions:
\begin{equation}
	\begin{split}
		k_{x,y,z} &\rightarrow \sin(k_{x,y,z}a)/a, \\
		k_{x,y,z}^2 &\rightarrow 2[1-\cos(k_{x,y,z}a)]/a^2.
	\end{split}
\end{equation}
Note that the substitution of the linear terms add artificial gap closings at $k_{x,y,z} = \pm \pi/a$, being removed again by the quadratic terms.
Integrating over the reciprocal space yields the following tight-binding Hamiltonian in real space consisting of on-site and nearest-neighbor hopping terms:
\begin{widetext}
\begin{equation}
	\begin{split}
		&\mathcal{H}_\lattice^\Zhang (\veck) = \sum_{i, \sigma} \left\{ \lef C_0^+ - 2 \frac{2 C_\perp^+ + C_z^+}{a^2} \rig e^\dag_{i \, \sigma} e_{i \, \sigma}
			+ \lef C_0^- - 2 \frac{2 C_\perp^- + C_z^-}{a^2} \rig h^\dag_{i \, \sigma} h_{i \, \sigma} \right\} \\
		&\; + \sum_{i, \sigma} \left\{ \frac{C_\perp^+}{a^2} ( e^\dag_{i + \hat{x} \, \sigma} e_{i \, \sigma} + e^\dag_{i + \hat{y} \, \sigma} e_{i \, \sigma})
			+ \frac{C_\perp^-}{a^2} (h^\dag_{i + \hat{x} \, \sigma} h_{i \, \sigma} + h^\dag_{i + \hat{y} \, \sigma} h_{i \, \sigma} )
			+ \frac{C_z^+}{a^2} e^\dag_{i + \hat{z} \, \sigma} e_{i \, \sigma}
			+ \frac{C_z^-}{a^2} h^\dag_{i + \hat{z} \, \sigma} h_{i \, \sigma} + \mathrm{h.c.} \right\} \\
		&\; + \sum_i \left\{ \frac{A_\perp}{2 a} ( - \imu e^\dag_{i + \hat{x} \, \uparrow} h_{i \, \downarrow} + \imu e^\dag_{i - \hat{x} \, \uparrow} h_{i \, \downarrow}
			- \imu e^\dag_{i + \hat{x} \, \downarrow} h_{i \, \uparrow} + \imu e^\dag_{i - \hat{x} \, \downarrow} h_{i \, \uparrow}
			- e^\dag_{i + \hat{y} \, \uparrow} h_{i \, \downarrow} + e^\dag_{i - \hat{y} \, \uparrow} h_{i \, \downarrow}
			+ e^\dag_{i + \hat{y} \, \downarrow} h_{i \, \uparrow} - e^\dag_{i - \hat{y} \, \downarrow} h_{i \, \uparrow} ) \right. \\
		&\qquad \quad \left.
			+ \frac{A_z}{2a} (- \imu e^\dag_{i + \hat{z} \, \uparrow} h_{i \, \uparrow} + \imu e^\dag_{i - \hat{z} \, \uparrow} h_{i \, \uparrow}
			+ \imu e^\dag_{i + \hat{z} \, \downarrow} h_{i \, \downarrow} - \imu e^\dag_{i - \hat{z} \, \downarrow} h_{i \, \downarrow}  ) + \mathrm{H.c.} \right\}, \qquad C_{0, \perp, z}^\pm \equiv C_{0, \perp, z} \pm M_{0, \perp, z},
	\end{split}
\end{equation}
\end{widetext}
with $e^\dagger$ and $h^\dagger$ creation operators for the E and H orbitals respectively, $i$ a summation index for the square lattice sites, and $\sigma$ summing over the spin degree of freedom. This Hamiltonian can directly be implemented in Kwant. The resulting band structure approaches that of the continuous model in the limit $a \rightarrow 0$ and forms a good approximation when $|\veck|<a^{-1}$.

\begin{acknowledgments}

The authors thank Giacomo Dolcetto, Wim Magnus and Bart Sor\'ee for fruitful discussions and acknowledge the support by the National Research Fund Luxembourg (ATTRACT Grant No.~7556175).

\end{acknowledgments}

\bibliographystyle{apsrev4-2}
\bibliography{2018_Moors_Magnetotransport_corr}

\end{document}